\documentclass[journal]{styles/IEEEtran}
\pdfoutput=1
\usepackage{amsmath,amssymb,epsfig,theorem,url,setspace,verbatim,multirow}
\usepackage{color}
\definecolor{myblue}{rgb}{0.2,0.4,1.00}

\newtheorem{thm}{Theorem}
\newtheorem{lem}{Lemma}

\newtheorem{example}{Example}
\newcommand{\refeq}[1]{\ensuremath{\stackrel{(#1)}{=}}}

\newcommand{\Perp}{\perp \! \! \! \perp}

\newcommand{\bfB}{\mathbf{B}}

\newcommand{\bfS}{\mathbf{S}}
\newcommand{\bfs}{\mathbf{s}}
\newcommand{\bfK}{\mathbf{K}}
\newcommand{\bfk}{\mathbf{k}}
\newcommand{\bfA}{\mathbf{A}}
\newcommand{\bfa}{\mathbf{a}}
\newcommand{\bfC}{\mathbf{C}}

\newcommand{\bfJ}{\mathbf{J}}

\newcommand{\bfD}{\mathbf{D}}

\newcommand{\bfH}{\mathbf H}
\newcommand{\bfM}{\mathbf M}
\newcommand{\cV}{{\cal V}}

\begin{document}

\title{A Theoretical Analysis of Authentication, Privacy and Reusability Across Secure Biometric Systems}

\author{ Ye~Wang, Shantanu Rane,~\emph{Member, IEEE},
  Stark~C.~Draper,~\emph{Member, IEEE}, Prakash
  Ishwar,~\emph{Senior Member, IEEE}

\thanks{Y.~Wang is with the Dept.~of Electrical and Computer
  Engineering, Boston University, Boston, MA 02215 (ye@bu.edu).}

\thanks{S.~Rane is with Mitsubishi Electric Research Laboratories
  (MERL), Cambridge, MA 02139 (rane@merl.com).}

\thanks{S.~C.~Draper is with the Dept.~of Electrical and Computer
  Engineering, University of Wisconsin, Madison, WI 53706
  (sdraper@ece.wisc.edu).}

\thanks{P.~Ishwar is with the Dept.~of Electrical and Computer
  Engineering, Boston University, Boston, MA 02215 (pi@bu.edu).}}

\maketitle

\begin{abstract}

We present a theoretical framework for the analysis of privacy and
security tradeoffs in secure biometric authentication systems.  We use
this framework to conduct a comparative information-theoretic analysis
of two biometric systems that are based on linear error correction
codes, namely fuzzy commitment and secure sketches.  We derive upper
bounds for the probability of false rejection ($P_{FR}$) and false
acceptance ($P_{FA}$) for these systems.  We use mutual information to
quantify the information leaked about a user's biometric identity, in
the scenario where one or multiple biometric enrollments of the user
are fully or partially compromised.  We also quantify the probability
of successful attack ($P_{SA}$) based on the compromised
information. Our analysis reveals that fuzzy commitment and secure
sketch systems have identical $P_{FR}, P_{FA}, P_{SA}$ and information
leakage, but secure sketch systems have lower storage requirements. We
analyze both single-factor (keyless) and two-factor (key-based)
variants of secure biometrics, and consider the most general scenarios
in which a single user may provide noisy biometric enrollments at
several access control devices, some of which may be subsequently
compromised by an attacker. Our analysis highlights the revocability
and reusability properties of key-based systems and exposes a subtle
design tradeoff between reducing information leakage from compromised
systems and preventing successful attacks on systems whose data have
not been compromised.
\end{abstract}


\section{Introduction}
\label{sec:introduction}

Human biometric measurements such as fingerprints, iris scans, face
images and ECG signals are attractive tools for identifying and
authenticating users in access control situations. Unlike conventional
identifying documents, biometrics are difficult to forge.  Unlike
passwords traditionally used for access control, they do not have to
be remembered. However, biometrics also present some new challenges
that are not encountered in traditional methods. Noise is a
characteristic feature of all biometric measurements; every
measurement is slightly different from all others.  In access control
systems, the issue of noise in biometric measurements is currently
tackled through pattern recognition. Specifically, a measurement of
the biometric is taken at the time of enrollment and stored in a
database of enrolled identities.  During authentication, the person in
question provides a ``test'' or a ``probe'' biometric for comparison
with the stored enrollment biometric.  If the probe and enrollment
biometric are sufficiently close according to a similarity metric
defined by the pattern recognition algorithm, then access is allowed.

Unfortunately, the standard method described above has a serious
drawback: an adversary who compromises the device gains access to the
enrollment biometric.  This is a major security hazard; the attacker
can subsequently use the enrollment biometric to gain repeated access
to the system, and to any other biometric-based systems in which the
user has enrolled.  This is also a privacy hazard; the attacker has
gained access to the user's identifying information and can henceforth
impersonate the user illegally. The seriousness of this hazard is
greatly increased by the fact that biometrics are inherent properties
of the human body and cannot be revoked and then re-issued like new
credit card numbers.  To mitigate growing concerns about security
hazards and identity theft, new approaches to biometrics have been
studied with a three-fold goal.  First, the data stored on the access
control device should provide little or no information about the
actual biometric.  Second, the stored data should not allow an
attacker to gain unauthorized access to the system or to impersonate
the identity of a legitimate user successfully.  Third, if the user's
stored data is known to have been compromised, then it should be
possible to revoke it and issue new stored data.  This should prevent
the adversary from gaining access or stealing the user's identity in
the future.

Secure biometric schemes proposed to fulfill the above desiderata fall
under one of two related categories, viz., {\em fuzzy
  commitment}~\cite{davida98ssp,juels99fuzzy,juels02isit,tuyls05avbpa,dodis04eurocrypt}
and {\em secure sketch}
schemes~\cite{dodis04eurocrypt,li06asiacrypt,sutcu07tifs,draper07ita,sutcu08isit}.
In fuzzy commitment a secret vector is combined with the user's
enrollment biometric via a {\em commitment function}. The output of
the commitment function is stored on the access control device.
Access control is accomplished by means of a {\em decommitment}
function.  The decommitment function takes as its inputs the stored
data and the user's probe biometric and attempts to recover the secret
vector.  If recovery is successful, access is allowed.  In contrast,
in secure sketch the user provides their biometric at enrollment and a
``sketch'' signal is derived and stored on the access control
device. When combined with a probe biometric from the legitimate user,
the enrollment biometric can be recovered. If the enrollment biometric
is recovered successfully, then access is allowed.  We later discuss
how to verify the correctness of this recovery or of successful
decommitment in fuzzy commitment.  Linear error correcting codes (ECC)
are the most widely used tool for constructing both fuzzy commitment
schemes~\cite{juels99fuzzy,juels02isit,nagar08icpr,nandakumar07tifs}
and secure sketch-based schemes~\cite{draper07ita,sutcu08isit}.

The relationship between secure sketches and fuzzy extractors was
examined in~\cite{dodis04eurocrypt} where it was shown that a secure
sketch implies the existence of a fuzzy extractor. In the present
paper, we analyze explicit ECC-based constructions of fuzzy commitment
and secure sketch.  We study both the security and privacy hazards
mentioned above.  Regarding the former, we derive upper bounds on the
false rejection rate (FRR) and false acceptance rate (FAR) for both
types of systems.  Regarding the latter, we characterize the
privacy leakage as the mutual
information between the compromised stored data and the user's
biometric. Further, a smart adversary may be able to increase their
likelihood of gaining access to a system above the FAR if they have
access to some partial compromise of stored data and condition their
attack on that knowledge.  We term this the probability of a
``successful attack'' ($P_{SA}$) and quantify it in some situations.
Our analysis establishes a strong statement of equivalence: secure
sketches and fuzzy commitment schemes are equivalent in terms of the
FRR, FAR, information leakage, and $P_{SA}$.

There have been many insightful studies of the information leakage
that occurs when data stored on the access control device is
compromised~\cite{boyen04ccs,dodis04eurocrypt,simoens09oakland,linnartz03avbpa}.
An important insight is that a useful sketch, i.e, one that correctly
authenticates noisy samples from a legitimate user, must leak some
information about the underlying
biometric~\cite{dodis04eurocrypt}. Extending this
idea,~\cite{boyen04ccs} considers a generalized challenge-response setting in
which a strong adversary examines sketches from several chosen
perturbations of the challenger's biometric, until the biometric has
been guessed completely.  We
consider a different scenario in which an adversary compromises a
chosen subset of the \emph{available} access control devices and,
knowing the error correcting codes associated with each, attempts to
attack the user's system.  
%Compared with~\cite{boyen04ccs}, our
%adversary is weaker in the sense that he can only choose from a fixed
%set of perturbations (the perturbations here result from the
%variability in the enrollment biometrics across the systems) rather
%than from an arbitrary set of perturbations. 
We think that this problem formulation is more reflective of emerging networks of biometric
systems.  Further, it raises many interesting challenges, e.g., we may
ask how to choose the perturbations or error correcting codes so as to
leak the least information about the user's biometric. In this sense,
our work is related to the privacy analysis
of~\cite{simoens09oakland}, where the authors consider a sketch
indistinguishability game and sketch irreversibility game and give
conditions on the ECC design that minimizes the adversary's
advantage. We note that, in the analyses
of~\cite{boyen04ccs,dodis04eurocrypt,simoens09oakland}, the emphasis
is on information leakage about the user's biometric as the
adversary's prime objective. In practice, however, the adversary may
have a second objective, namely to compromise some devices and use the
information gained to login to other devices.  It may not be necessary
to discover the user's biometric.  Our analysis reveals a subtle
conflict between reducing information leakage from compromised systems
and preventing successful attacks on systems whose data have not been
compromised.

%As we shall see, this is pertinent in the case of two-factor secure biometric schemes,
%in which, unlike one-factor schemes, a compromised sketch can be designed to leak 
%zero information about the biometric. 

A different, but related, line of work focuses on the problem of
secret key agreement via public discussion~\cite{maurer93it,
  ahlswede93IT, ignatenko09tifs, lai11tifs1, lai11tifs2}. In this problem two
parties hold correlated pieces of information and desire to generate
matching secret keys through a public discussion. However, an
eavesdropper who taps into the public discussion should learn nothing
about the keys.  Of interest in this line of work is the fundamental
{\em asymptotic} tradeoff between the secret key rate (security) and
biometric information leakage (privacy).  Secret key agreement by
itself does not form a biometric authentication system but it can be
used to construct one. In contrast, we explicitly analyze the
fundamental {\em non-asymptotic} privacy-security tradeoff in
biometric systems that are based on linear ECCs and explicitly relate
them to ECC-design parameters.

The remainder of this paper is organized as follows:
Section~\ref{sec:framework} describes a general framework for
analyzing secure biometrics and defines the metrics by which security
and privacy\footnote{In this work, compromising privacy refers to
  leaking information about the user's biometric, while compromising
  security refers to gaining access to the system.} are evaluated. In
Section~\ref{sec:Systems}, we describe how to realize fuzzy commitment
and secure sketch schemes using linear ECCs.  We show the
equivalence between the realizations of fuzzy commitment and secure
sketch in terms of their security and privacy metrics. In
Section~\ref{sec:revocability}, we expand our attention to include
multiple devices. We derive the information leakage for attack
scenarios in which an adversary compromises the stored data and/or
secret keys of multiple devices.  We show how the information leakage
depends on the ECCs used at the devices.We characterize how the selection of the
ECCs affects the probability that the adversary can use information
gained from the compromised devices to successfully attack (i.e., gain
access to) uncompromised devices, and how this objective
conflicts with the aim of minimizing information leaked about the
user's biometric. Section~\ref{sec:conclusions} concludes the paper.

\begin{figure}
\centering
\includegraphics[width=3.4in]{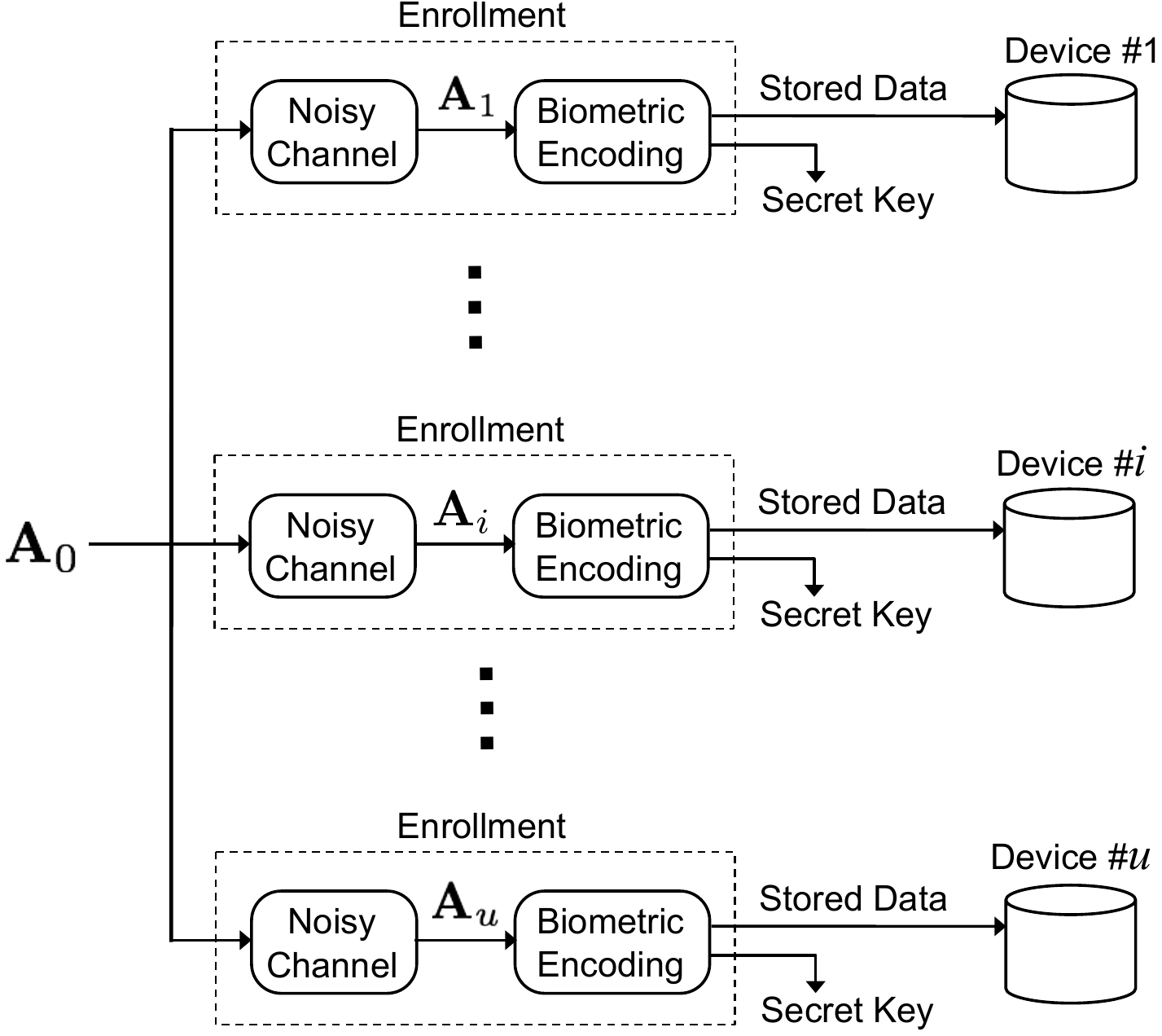}
\caption{Noisy measurements ${\bf A}_1, \ldots, {\bf A}_u$ of a user's
  underlying biometric ${\bf A}_0$ are encoded at each access control
  device to generate authentication data, which is stored in the
  device, and a secret key.  Our goal is to analyze the tradeoffs
  between authentication performance and information leakage from
  compromised stored authentication data and secret keys. }
\label{fig:bioceo}
\end{figure}

\section{A Generalized Secure Biometrics Framework}
\label{sec:framework}

Consider the scenario with several access control devices shown in
Fig.~\ref{fig:bioceo}.  A user has a biometric $\mathbf{A}_0$ given by
nature. He enrolls at several access control devices using noisy
measurements $\mathbf{A}_i$ of the underlying biometric
$\mathbf{A}_0$. From each measurement $\mathbf{A}_i$, encoded data is
extracted and stored on the respective device to aid in
authentication. Optionally, a secret key or password is provided to
the user. A legitimate user should be able to gain access to any of
the devices by providing a probe biometric that is again a noisy
measurement of the underlying $\mathbf{A}_0$. Any analysis of the
privacy and security tradeoffs in secure biometrics must take into
account not only the authentication performance but also the
information leakage when the stored data and/or keys for one or more
devices are compromised.  

With the above motivation, we start by presenting an abstract model of
a secure biometric system for a single access control device in
Section~\ref{subsec:model}.  We then describe design objectives in
terms of the system's performance metrics in
Section~\ref{subsec:metrics}.  
%We characterize the objectives using an
%information theoretic analysis in Section~\ref{sec:Systems}. Analysis
%of issues and performance across multiple access control devices is
%presented in Section~\ref{sec:MultSys}.

\begin{comment}
We first describe a model for single biometric system in isolation.
This allows us to define and analyze the basic security requirements
of such a system.  However, later in Section~\ref{sec:MultSys}, we
consider multiple, parallel systems in order to analyze the more
complex security requirements of resistance to linkage attacks and
revocability.
\end{comment}
 
\subsection{Model of a Secure Biometric System}
\label{subsec:model}

\begin{figure}[!htb]
\centering
\includegraphics[width=3.4in]{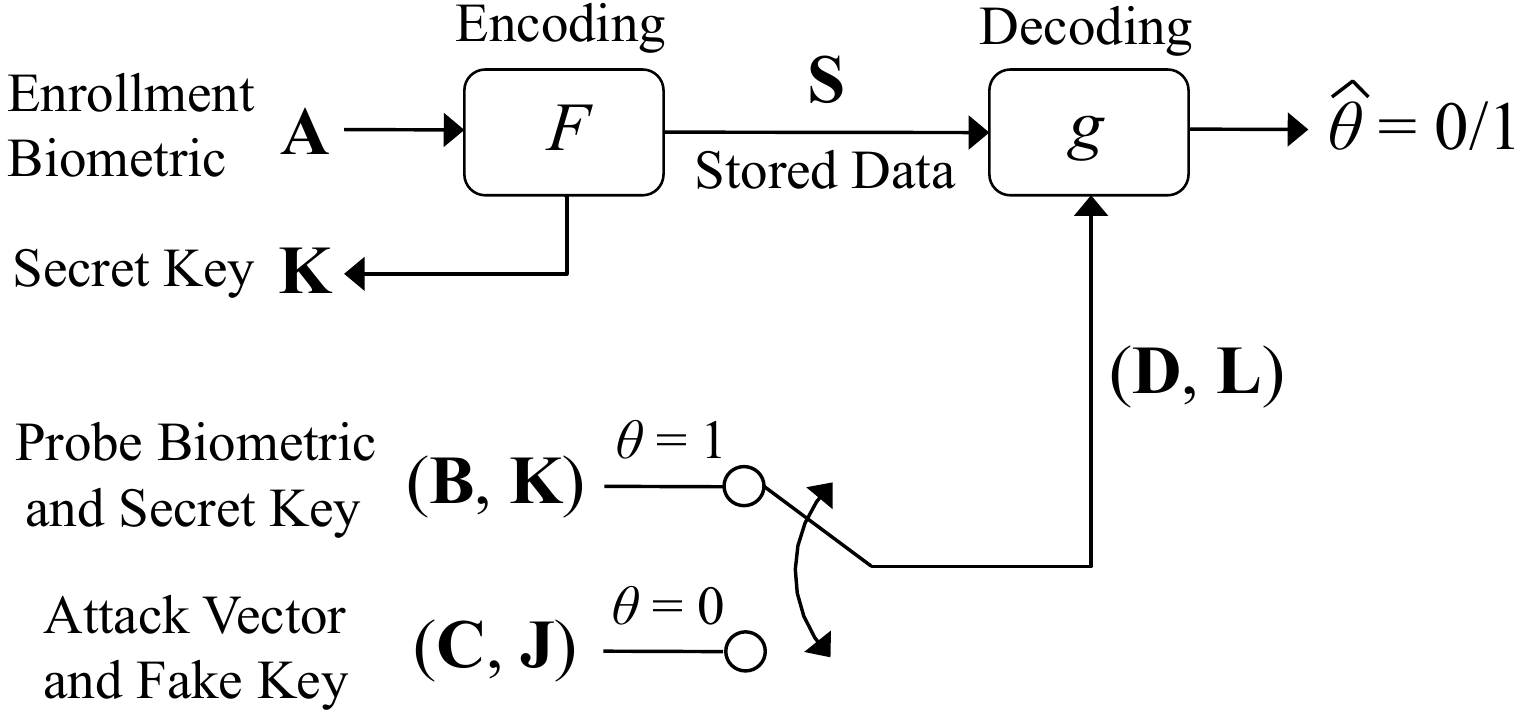}
\caption{Generalized model of a secure biometric system for a single
  access control device. This model encompasses both fuzzy
  commitment-based and secure sketch-based realizations that are
  described and analyzed in Section~\ref{sec:Systems}. For keyless
  realizations, $\mathbf{K}$ is null. For two-factor realizations,
  $\mathbf{K}$ is a secret key output by the randomized encoding
  function. Given the probe biometric and, in two-factor realizations,
  a secret key, the decoder solves a hypothesis testing problem to
  determine if the user is genuine or an impostor.}
\label{fig:framework} 
\end{figure}
Figure~\ref{fig:framework} depicts a generalized model of a secure
biometric system for a single access control device. The system
consists of encoding and decoding modules that manipulate features
extracted from measurements of human biometrics. In biometrics
parlance the terms ``biometric'', ``biometric measurement'', and
``biometric feature vector'', have different meanings. A fingerprint,
iris, or a face is a {\em biometric}, the {\em measurement} of which
produces a digitized image from which {\em features} are extracted for
authentication or recognition. However, for brevity of exposition, we
interchangeably use the terms ``biometric'' and ``biometric measurement'' to
denote a biometric feature vector. We make the additional simplifying
assumption that all feature vectors and secret keys are length-$n$
sequences of binary numbers.  The generalization to non-binary finite
alphabets is straightforward.

{\bf Biometric Measurement Model:} The process of measuring a
biometric, extracting suitable feature vectors, and converting them to
length-$n$ binary sequences is inherently prone to sensing
uncertainty, e.g., in the orientation, size, and illumination of an
iris or a face, as well as noise in the sensing elements. Since we are
interested in scenarios where a user can enroll the same biometric at
multiple access control devices (see Section~\ref{sec:MultSys}), we
posit an underlying ``ground truth'' length-$n$ binary biometric
feature vector $\mathbf{A}_0 := (A_{0,1},\ldots,A_{0,n})$ whose components have an
i.i.d. Bernoulli($0.5$) distribution.\footnote{Binarized feature vectors extracted
  from biometric measurements are, in general, neither independent nor
  identically distributed. It is, however, possible to design feature
  transformation algorithms that can convert them into binary feature
  vectors the statistics of which are quite close to those of
  i.i.d. Bernoulli($0.5$) bits~\cite{sutcu08isit}.} We need to model
the combined effect of a measurement followed by the extraction of a
length-$n$ binarized feature vector (or, for brevity, the biometric
measurement).  We model this as component-wise modulo-two addition of
$\mathbf{A}_0$ with a length-$n$ i.i.d.  Bernoulli ``noise''
sequence.  The noise sequence is assumed to be independent of the
ground truth and any previous and future measurement noise
sequences. In the language of information theory, the biometric
measurement is the output of a ``binary symmetric channel'' (BSC),  
%with crossover probability $p$
where the channel input is
$\mathbf{A}_0$. Thus, at enrollment, the user provides an enrollment
biometric measurement $\mathbf{A} := (A_1,\ldots,A_n)$ which 
is the output of a BSC with crossover probability $p_1$ and channel input
$\mathbf{A}_0$.  Similarly, at
authentication, the user provides a probe biometric measurement
$\mathbf{B} := (B_1,\ldots,B_n)$, which is the output of a BSC with
crossover probability $\alpha$ and channel input $\mathbf{A}_0$. 
This second probe measurement is
used by the decoding module of the access control device to verify the
user's identity.  We further assume that $p_1 \in [0,0.5)$ and $\alpha
\in [0,0.5)$, i.e., it is more likely that coordinates of the biometric
measurement and probe measurement match than that they do not.
To see the statistical dependency between $\mathbf{A}$ and $\mathbf{B}$,
observe that $\mathbf{A}_0$, $\mathbf{A}$ and $\mathbf{B}$ are all
i.i.d Bernoulli-0.5 sequences. This, along with the BSC channel 
dependency explained above, means that $\mathbf{A}$ and $\mathbf{B}$
are, in turn, related by a BSC with crossover probability 
$p=p_1 * \alpha = p_1(1 - \alpha) + (1- p_1) \alpha$.
%Therefore, the elements $(A_i, B_i)$ are i.i.d. samples of a doubly
%symmetric binary source with crossover probability $p = p_1(1-p_2) +
%p_2(1-p_1) < 0.5$.

{\bf Enrollment:} The (potentially randomized) encoding function
$F(\cdot)$ takes the enrollment biometric $\mathbf{A}$ as input and
produces as outputs $\mathbf{S} \in \mathcal{S}$, $|\mathcal{S}| <
\infty$, which is stored on the access control device.  Optionally, a
key vector $\mathbf{K} \in \mathcal{K}$, $|\mathcal{K}| < \infty$,
which is returned to the user, is also produced. Thus,
$(\mathbf{S},\mathbf{K}) = F(\mathbf{A})$. The encoding function is
governed by the conditional distribution
$P_{\mathbf{S},\mathbf{K}|\mathbf{A}}$. Depending upon the physical
realization of the system, the user may be required to carry the key
$\mathbf{K}$ on a smart card. Such systems are called {\em two-factor}
systems because both the key and the stored data are needed for
authentication. Systems where $\mathbf{K}$ is null are called {\em
  keyless} systems; they do not require the use of a smart card.

{\bf Authentication:} To perform biometric authentication, a
legitimate user provides the probe biometric $\mathbf{B}$ and the key
$\mathbf{K}$.  An adversary, on the other hand, provides a stolen or
artificially synthesized biometric $\mathbf{C}$ and a stolen or
artificially synthesized key $\mathbf{J}$. The presence of the
legitimate user or the adversary is indicated by the unknown binary
parameter $\theta$.  Let $(\mathbf{D},\mathbf{L})$ denote the
(biometric, key) pair that is provided during the authentication step.
We write
\[
(\mathbf{D},\mathbf{L}) := \begin{cases}
(\mathbf{B},\mathbf{K}), & \text{if } \theta = 1,\\
(\mathbf{C},\mathbf{J}), & \text{if } \theta = 0.
\end{cases}
\]
The authentication decision is computed by the decoding function as
$\hat{\theta} = g(\mathbf{D},\mathbf{L},\mathbf{S})$. In keyless
systems, the procedure is similar with $\mathbf{K}$, $\mathbf{J}$, and
$\mathbf{L}$ removed from the above description.

\subsection{Performance Metrics}
\label{subsec:metrics}
We now define metrics used to evaluate the performance of the secure
biometric system of Fig.~\ref{fig:framework}. For example, it is
necessary to quantify how reliably the system authenticates a
genuine user and rejects an impostor, to quantify how much information
is leaked about the underlying biometric when the stored data and/or
the secret key are compromised, and so on.
\begin{enumerate}
\item \textbf{Probability of Missed Detection}: This quantity is also
  called the False Rejection Rate (FRR), defined as
\[
P_{FR} := \Pr \big[\hat{\theta} = 0 | \theta = 1 \big] = \Pr \big[ g(\mathbf{B},\mathbf{K},\mathbf{S})=0 \big] .
\]
The $P_{FR}$ depends only on the known statistics of
$(\mathbf{A},\mathbf{B},\mathbf{K})$ and the specification of the
system, $F(\cdot)$ and $g(\cdot)$. A low value of $P_{FR}$ indicates
that the system reliably authenticates a genuine user.  Thus $P_{FR}$
quantifies the accuracy of the biometric system.

\item \textbf{Probability of False Detection}: A {\em baseline}
  probability of false detection, also called the False Acceptance
  Rate (FAR) is the worst-case probability of false detection across
  all attack vectors and keys that can be generated without any
  knowledge of the ground truth or of any measurements, keys, or
  stored data.  It is defined as
\begin{align}
P_{FA} &:= \max_{p_{\mathbf{C},\mathbf{J}}} \Pr \big[\hat{\theta} = 1 | \theta = 0 \big]  \notag\\
              &= \max_{p_{\mathbf{C},\mathbf{J}}}  \Pr \big[ g(\mathbf{C},\mathbf{J},\mathbf{S})=1 \big] , \notag
\end{align}
where $(\mathbf{C},\mathbf{J})$ is independent of
$(\mathbf{A}_0,\mathbf{A},\mathbf{B},\mathbf{K},\mathbf{S})$.
%This definition is very general; if cryptographic functions are used
%in the biometric system, then additional assumptions on the adversary
%(e.g., computationally bounded) can be captured by limiting the class
%of admissible distributions to be maximized over.  
A low value of $P_{FA}$ indicates that the system reliably prevents
impostors from gaining access to the system by pure chance. Thus
$P_{FA}$ quantifies one aspect of the security of the biometric
system. Typically, a system designer is faced with choosing an
appropriate tradeoff between $P_{FA}$ and $P_{FR}$.

\item \textbf{Privacy Leakage}: We measure the information leaked
  about the enrollment biometric $\mathbf{A}$ (respectively the
  ground truth $\mathbf{A}_0$) in various scenarios of data exposure.
  These include when either the stored data $\mathbf{S}$, the secret
  key $\mathbf{K}$, or both are compromised.  We characterize the
  various scenarios using the following mutual information
  quantities: $I(\mathbf{A};\mathbf{S})$, $I(\mathbf{A};\mathbf{K})$,
  and $I(\mathbf{A};\mathbf{S},\mathbf{K})$ (respectively
  $I(\mathbf{A}_0;\mathbf{S})$, $I(\mathbf{A}_0;\mathbf{K})$, and
  $I(\mathbf{A}_0;\mathbf{S},\mathbf{K})$).  These are
  information-theoretic measures of independence.\footnote{Mutual
    information between two sets of quantities is always non-negative
    and is equal to zero if, and only if, the two sets are independent
    \cite{Gallager-1968}.  Furthermore, we can always write the mutual
    information between two random quantities $\mathbf{X}$ and
    $\mathbf{Y}$ as $I(\mathbf{X}; \mathbf{Y}) = H(\mathbf{X}) -
    H(\mathbf{X} | \mathbf{Y})$ where $H(\cdot)$ and $H(\cdot|\cdot)$
    are, respectively, the entropy and conditional entropy of the
    argument(s).  Thus, mutual information characterizes the {\em
      reduction in uncertainty} about one random quantity, $\mathbf{X}$,
    when given knowledge of another, $\mathbf{Y}$.}
%  Since the biometric is an inherent personal property of the user,
%  these measures quantify the privacy leakage in an information
%  theoretic sense.
 
\item \textbf{Probability of Successful Attack}: In the event of data
  exposure, the probability of false detection could increase beyond
  the nominal value of $P_{FA}$. In addition to exposure of the stored
  data $\mathbf{S}$ and the secret key $\mathbf{K}$, mentioned above,
  we may also need to consider scenarios where an adversary coercively
  gains access to $\mathbf{A}$ as well. We need to capture the
  possibility that the attacker's biometric-key pair
  $(\mathbf{C},\mathbf{J})$ is generated using knowledge of the
  compromised data $\mathcal{V} \subseteq \{\mathbf{A}, \mathbf{S},
  \mathbf{K}\}$.  We denote by $P_{SA}$ the probability of false
  detection in such situations, defined as
\begin{align}
P_{SA}(\mathcal{V}) &:= \max_{p_{\mathbf{C},\mathbf{J}|\mathcal{V}}}
\Pr \big[ \hat{\theta} = 1 | \theta = 0 \big] \notag \\
     &= \max_{p_{\mathbf{C},\mathbf{J}|\mathcal{V}}}
 \Pr \big[ g(\mathbf{C},\mathbf{J},\mathbf{S})=1 \big] . \notag
\end{align}
We refer to $P_{SA}(\mathcal{V})$ as the ``Successful Attack Rate''
(SAR) to distinguish it from $P_{FA}$. The SAR captures the
probability of false detection when an adversary's attack is aided
by knowledge of $\mathcal{V}$. We note, in passing, that in any
keyless or two-factor system, knowledge of the stored data
$\mathbf{S}$ can drastically improve the ability of the adversary to
gain access, thus compromising the security of the system. We will
characterize this effect in Theorem~\ref{thm:SARgivenS}. 
Ideally, in two-factor systems, if an attacker has knowledge of only
one factor --- i.e., either the enrollment biometric $\mathbf{A}$ or
the key $\mathbf{K}$, but not both --- they will not be able to use
that information to improve their ability to authenticate falsely.
This motivates the following definition. We say that a system is {\em
  two-factor secure} if $P_{SA}(\mathbf{A}) = P_{SA}(\mathbf{K}) =
P_{FA}$.

\item \textbf{Storage Requirements}: Lastly, the system data storage
  requirement is given by the minimum number of bits needed to
  represent $\mathbf{S}$.  This is not more than $\log_2
  |\mathcal{S}|$ bits. The key length requirement is given by the
  minimum number of bits needed to represent $\mathbf{K}$, which is
  not more than $\log_2 |\mathcal{K}|$ bits.

\end{enumerate}

\section{System Constructions}
\label{sec:Systems}

In this section, we discuss a single access control device 
\emph{in isolation}, and analyze system privacy and security.  We describe two types of
systems, the first is a fuzzy commitment system and the second is a secure
sketch system; for both, we assume an implementation based on linear error
correcting codes.  We detail both keyless and keyed (two-factor)
variants.  The linear error correcting code construction allows us to
demonstrate a number of performance-equivalence properties
between fuzzy commitment and secure sketch systems.  Considerations of
privacy and security for a network of access control devices is
deferred to Sec.~\ref{sec:revocability}.

\subsection{Fuzzy Commitment Systems based on ECC}
\label{sec:fuzzycommitment}

\begin{figure}[t]
\centering
\includegraphics[width=3.5in]{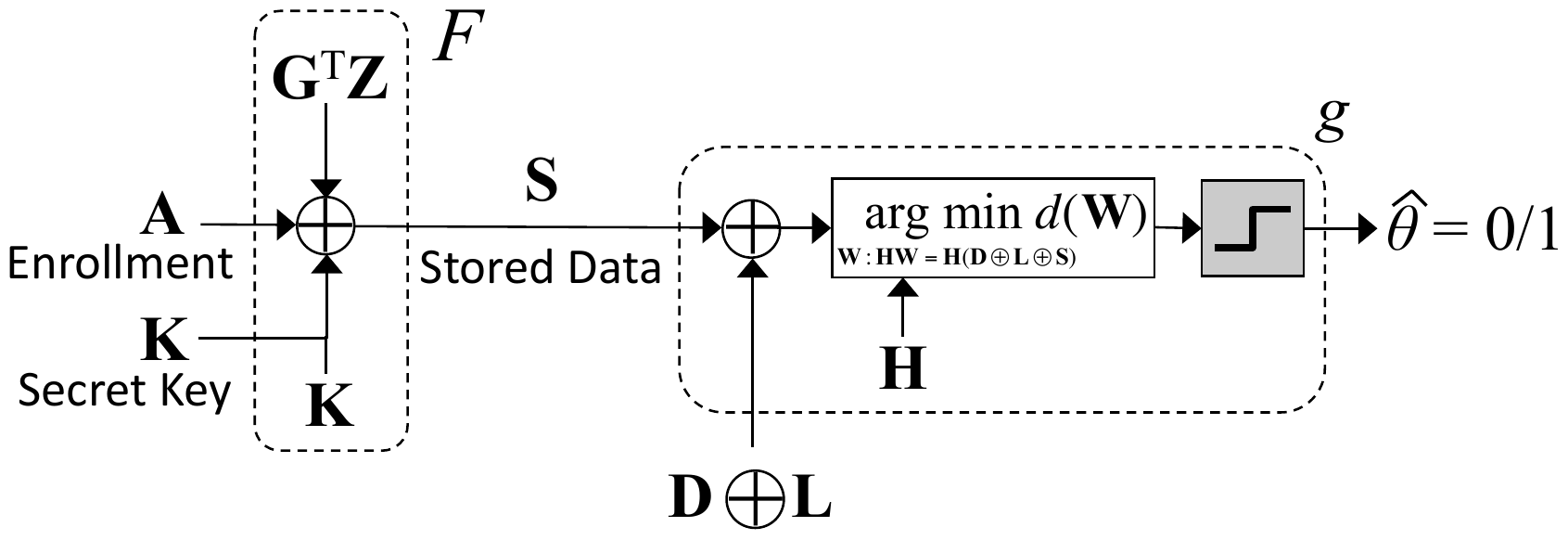}
\caption{A two-factor fuzzy commitment system stores the bitwise XOR
of a randomly generated codeword of a linear error correcting code,
the enrollment biometric, and a randomly generated secret key.}
\label{fig:fuzzycommitment2factor}
\end{figure}

\begin{figure}[t]
\centering
\includegraphics[width=3.5in]{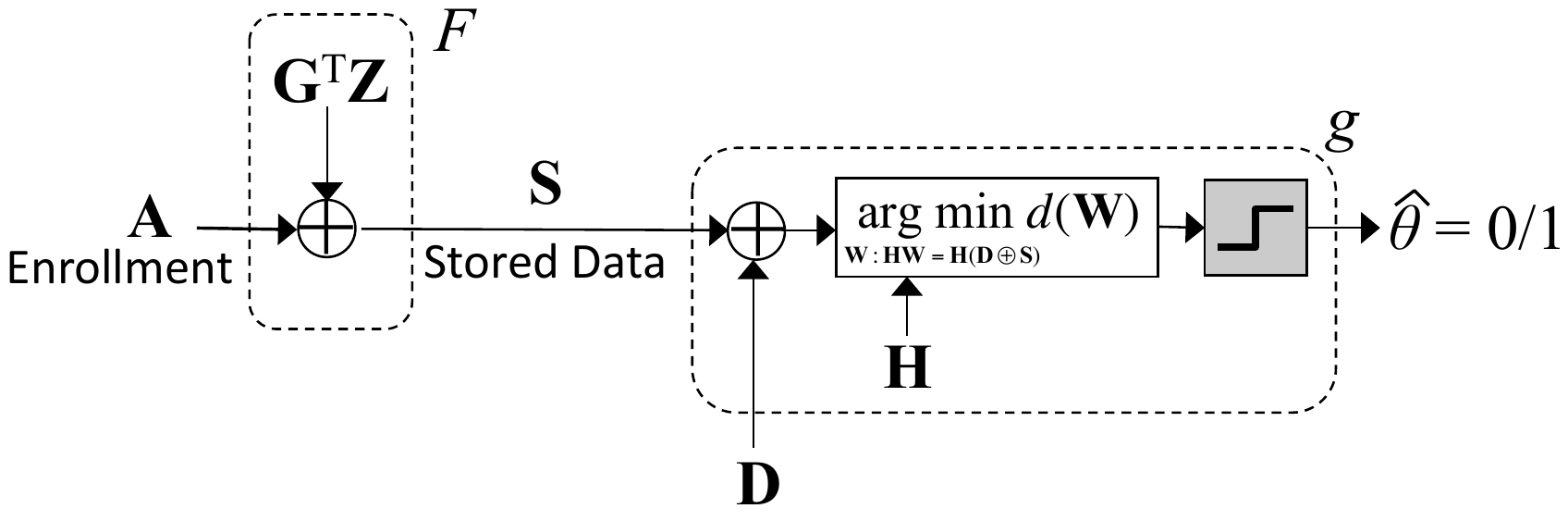}
\caption{A keyless fuzzy commitment system stores the bitwise XOR of a
randomly generated codeword of a linear error correcting code and the
enrollment biometric.}
\label{fig:fuzzycommitmentsimple}
\end{figure}

A {\em fuzzy commitment} scheme binds a random vector to an enrollment
biometric $\mathbf{A}$ to produce a length-$n$ stored data vector
$\mathbf{S}$. This is diagrammed in
Fig.~\ref{fig:fuzzycommitment2factor} for the case of a two-factor
(keyed) system.  The keyless variant, shown in
Fig.~\ref{fig:fuzzycommitmentsimple}, is the special case where 
the smart card key $\mathbf{K}$ and decoding key $\mathbf{L}$ are both
the all-zero sequence.  Note that both systems fit within the general
framework of Fig.~\ref{fig:framework}.

We exclusively consider fuzzy commitment schemes wherein the random
vector corresponds to a uniformly selected codeword of a binary
$[n,k]$ linear error correcting code.  We use $\mathbf{G}$ to denote
the code's $k \times n$ generator matrix and $\mathbf{H}$ to denote
the code's $m \times n$ parity check matrix with $m = n-k$.

{\bf Enrollment:} The enrollment procedure first generates two
independent i.i.d.\ Bernoulli$(0.5)$ sequences, the key sequence
$\mathbf{K} := (K_1,\ldots,K_n)$ and the auxiliary sequence,
$\mathbf{Z} := (Z_1,\ldots,Z_k)$. The auxiliary sequence $\mathbf{Z}$
selects a codeword $\mathbf{G}^\mathrm{T}\mathbf{Z}$ uniformly from
the set of all codewords of the linear error correction code with
generator matrix $\mathbf{G}$. The codeword is then additively
perturbed by the enrollment biometric $\mathbf{A}$ and the result is
additively masked by the randomly generated key sequence $\mathbf{K}$
to produce the stored data $\mathbf{S}$:
\[
\mathbf{S} = \mathbf{A} \oplus \mathbf{G}^\mathrm{T}\mathbf{Z} \oplus
\mathbf{K}.
\]

{\bf Authentication:} At authentication, the system has access to the
stored data $\mathbf{S}$ and is presented with the pair $(\mathbf{D},
\mathbf{L})$. The authentication procedure consists of two
steps. First, syndrome decoding is performed to produce an estimate
$\hat{\mathbf{W}}$ of the error vector $\mathbf{A} \oplus \mathbf{D}$
as follows:
\[
\hat{\mathbf{W}} = \mathop{\arg \min}_{\mathbf{W} :
\mathbf{H}\mathbf{W} = \mathbf{H}(\mathbf{D} \oplus \mathbf{L} \oplus
\mathbf{S})} d(\mathbf{W}),
\]
where $d(\cdot)$ is the Hamming weight.  If $\mathbf{L} = \mathbf{K}$,
the masking effect of the key is canceled out and the syndrome
decoding procedure is then operationally equivalent to the optimal
channel decoding of the codeword $\mathbf{G}^\mathrm{T}\mathbf{Z}$
when corrupted by $\mathbf{A} \oplus \mathbf{D}$.  Second, given
$\hat{\mathbf{W}}$, an estimate $\hat{\theta}$ of $\theta$ is made
as
\begin{equation}
d(\hat{\mathbf{W}}) \mathop{\lessgtr}^{\hat{\theta} = 1}_{\hat{\theta}
  = 0} \tau n. \label{eq.thresholdTest}
\end{equation}
If $\hat{\theta} = 1$ the decision is made that the biometric
$\mathbf{A}$ and the probe $\mathbf{D}$ are close enough (the estimate
of this distance is the weight of $\hat{\mathbf{W}}$) that access
should be granted.

We make the following assumptions about system operating parameters.
Recall that if $\mathbf{L} = \mathbf{K}$ the decoding process is the
same as optimal channel decoding.  This implies that if the rate of
the error correcting code (specified by the choice of $\mathbf{H}$) is
below the {\em channel capacity} of the binary symmetric channel (BSC)
with crossover probability $\tau$, BSC($\tau$), then the the estimate
$\hat{\mathbf{W}}$ will equal $\mathbf{A} \oplus \mathbf{D}$ with high
probability.  Our first assumption is thus that the rate
$R = k/n$ of the code $\mathbf{G}$ satisfies
\[
R = k/n < 1 - h_b(\tau),
\]
where $1 - h_b(\tau)$ is the BSC($\tau$) channel capacity and $h_b(p)
:= - p \log_2 p - (1-p) \log_2 (1-p)$ is the binary entropy function.
Second, we require $\tau$ to be larger than $p$ but smaller than
$0.5$, i.e., $0.5 > \tau > p$.  Recall that $p$ is the noise parameter
of the probe channel, the BSC($p$).  With this relation between $p$
and $\tau$ we write
\[
R = k/n < 1 - h_b(\tau) < 1 - h_b(p),
\]
or, equivalently, 
\[\frac{m}{n} > h_b(\tau) > h_b(p).
\]

In many practical realizations of fuzzy commitment the threshold
test~(\ref{eq.thresholdTest}) is replaced with a hash check.  Namely,
in order to verify whether the random vector $\mathbf{G}^T \mathbf{Z}$
has been recovered exactly, a cryptographic hash of $\mathbf{G}^T
\mathbf{Z}$ (alternately of $\mathbf{Z}$) is also stored at the access
control device. This stored hash must match the hash of the
$\mathbf{D} \oplus \mathbf{L} \oplus \mathbf{S} \oplus
\hat{\mathbf{W}}$ for access to be granted. However, cryptographic
hashes are not information theoretically secure, they are only
computationally secure.  Since our focus is on information
theoretic security, a cryptographic hash cannot be used as part of our
system.  Thus, in the systems analyzed in this work, we do not use
cryptographic hashes and, instead, rely on the threshold
test~(\ref{eq.thresholdTest}).

\begin{comment}
{\bf Enrollment:} The enrollment procedure first generates an
independent i.i.d.\ Bernoulli$(0.5)$ sequence $\mathbf{Z} :=
(Z_1,\ldots,Z_k)$. The stored data is then computed as
%
\[
\mathbf{S} = \mathbf{A} \oplus \mathbf{G}^\mathrm{T}\mathbf{Z},
\]
%
and the smart card key $\mathbf{K}$ is null.

{\bf Authentication:} The authentication procedure first performs
syndrome decoding to recover
%
\[
\hat{\mathbf{W}} = \mathop{\arg \min}_{\mathbf{W} :
  \mathbf{H}\mathbf{W} = \mathbf{H}(\mathbf{D} \oplus \mathbf{S})}
  d(\mathbf{W}),
\]
%
where $d(\cdot)$ is the Hamming weight. This is operationally
equivalent to the optimal channel decoding of
$\mathbf{G}^\mathrm{T}\mathbf{Z}$ corrupted by $\mathbf{A} \oplus
\mathbf{D}$, where $\hat{\mathbf{W}}$ is the corresponding optimal
recovery of $\mathbf{A} \oplus \mathbf{D}$. The authentication
decision is made via the following threshold test,
%
\[
\hat{\theta} = d(\hat{\mathbf{W}}) \mathop{\lessgtr}^1_0 \tau n,
\]
%
which accepts or rejects based on the closeness of the probe biometric
feature vector to the enrollment biometric feature vector.
\end{comment}

%
\begin{figure}[t]
\centering
\includegraphics[width=3.5in]{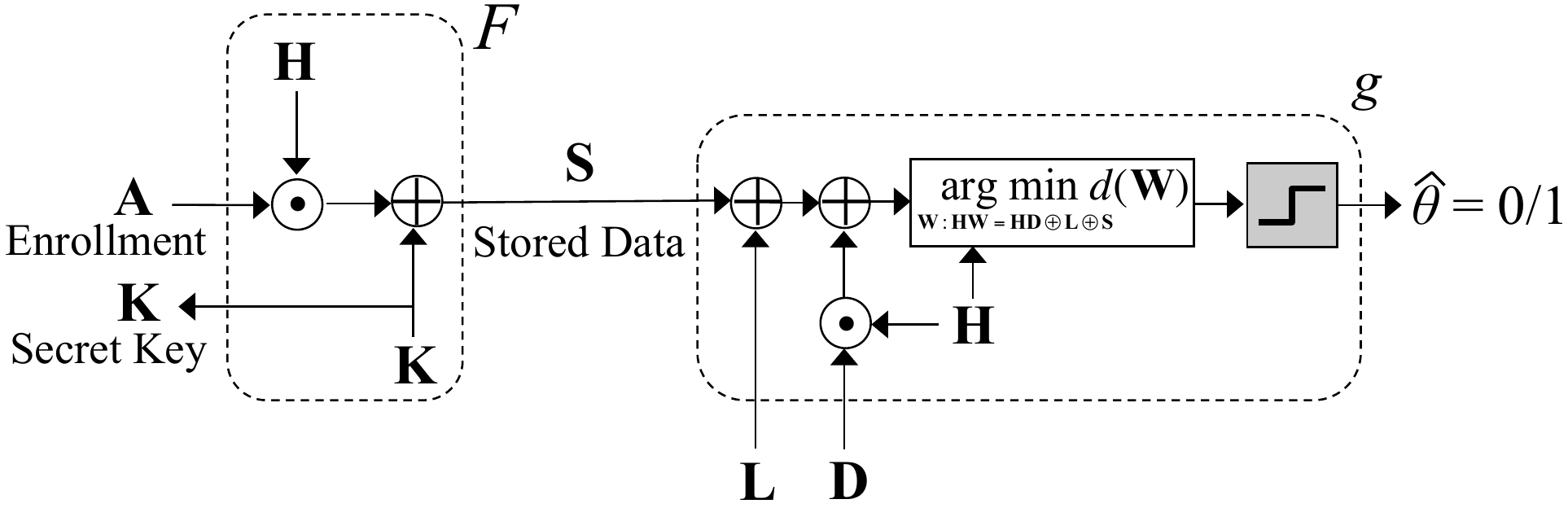}
\caption{A two-factor secure sketch system stores the bitwise XOR of
the syndrome vector of a linear error correcting code generated by the
enrollment biometric and a randomly generated secret key.}
\label{fig:securesketch2factor}
\end{figure}
\begin{figure}[t]
\centering
\includegraphics[width=3.5in]{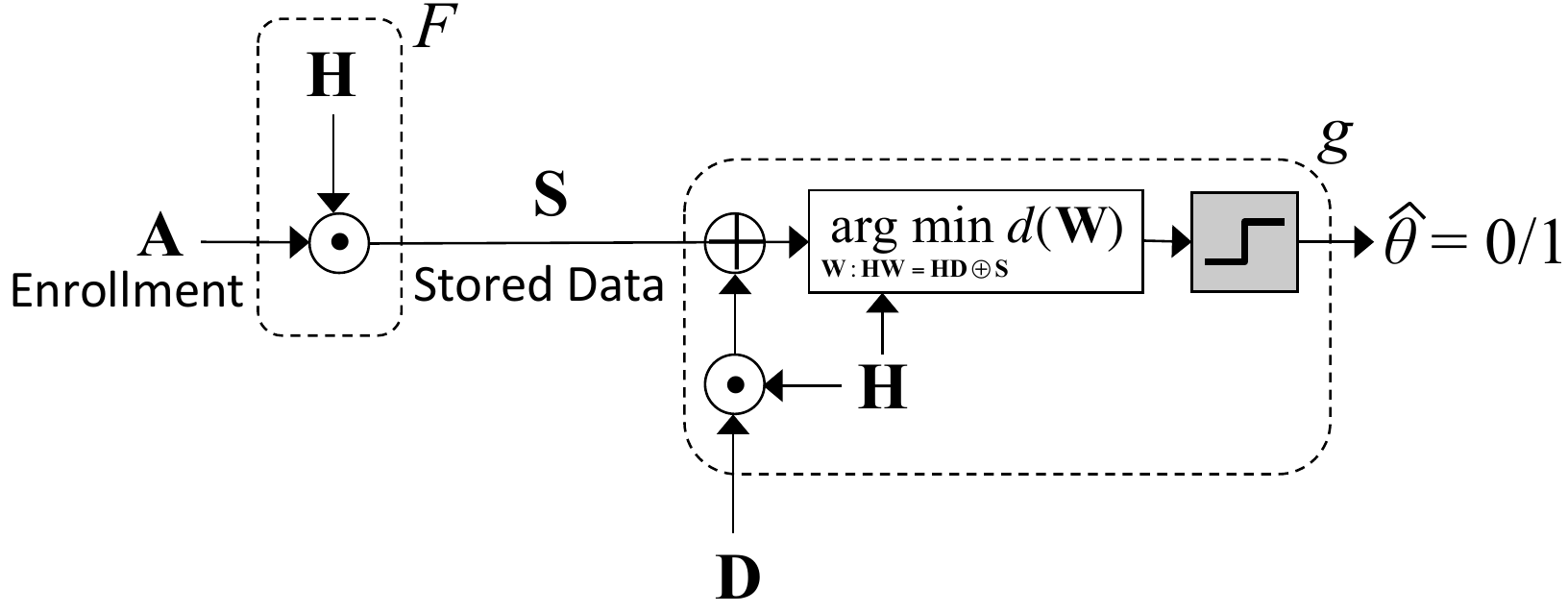}
\caption{A keyless secure sketch system stores the syndrome vector of
a linear error correcting code generated by the enrollment biometric.}
\label{fig:securesketchsimple}
\end{figure}

\subsection{Secure Sketch Systems based on ECC}
\label{sec:securesketch}

We now introduce the second family of biometric storage systems studied,
called {\em secure sketch} systems.  While, as was the case for fuzzy
commitment, there are other ways to develop a secure sketch, we
concentrate on secure sketches implemented using linear error
correcting codes.  The baseline two-factor secure sketch scheme is 
diagrammed in Fig.~\ref{fig:securesketch2factor} and the keyless
variant in Fig.~\ref{fig:securesketchsimple}.  Following the notation
of Sec.~\ref{sec:fuzzycommitment} we denote by $\mathbf{H}$ the $m
\times n$ parity check matrix of a binary $[n,k]$ linear error
correcting code with $m = n-k$.

{\bf Enrollment:} The enrollment procedure first generates the key
sequence $\mathbf{K} := (K_1,\ldots,K_m)$ as an independent i.i.d.\
Bernoulli$(0.5)$ sequence. The stored data $\mathbf{S}$ is the
length-$m$ syndrome $\mathbf{H} \mathbf{A}$ of enrollment biometric
feature vector masked by the key,
\[
\mathbf{S} = \mathbf{H}\mathbf{A} \oplus \mathbf{K}.
\]

{\bf Authentication:} The authentication procedure performs syndrome
decoding to produce an estimate $\hat{\mathbf{W}}$ of $\mathbf{A}
\oplus \mathbf{D}$ as
\[
\hat{\mathbf{W}} = \mathop{\arg \min}_{\mathbf{W}:
  \mathbf{H}\mathbf{W} = \mathbf{H}\mathbf{D} \oplus \mathbf{L} \oplus
  \mathbf{S}} d(\mathbf{W}).
\]
The authentication decision is made using threshold test
\[
d(\hat{\mathbf{W}}) \mathop{\lessgtr}^{\hat{\theta} = 1}_{\hat{\theta}
  = 0} \tau n.
\]
The assumptions on the values of $\tau$ and the coding rate $R$ of the
ECC are identical to those made in Sec.~\ref{sec:fuzzycommitment}.  In
practical implementations, cryptographic hashes are often also stored
and used to verify the correctness of the syndrome decoding procedure.
However, for the reasons already discussed in the context of fuzzy
commitment, we do not employ cryptographic hashes in our analysis.

\begin{comment}
{\bf Enrollment:} The enrollment procedure stores the syndrome of
enrollment biometric feature vector $\mathbf{A}$,
%
\[
\mathbf{S} = F(\mathbf{A}) = \mathbf{H}\mathbf{A},
\]
%
and the smart card key $\mathbf{K}$ is null.

{\bf Authentication:} The authentication procedure first performs
syndrome decoding to recover
%
\[
\hat{\mathbf{W}} = \mathop{\arg \min}_{\mathbf{W}: \mathbf{H}\mathbf{W} = \mathbf{H}\mathbf{D} \oplus \mathbf{S}} d(\mathbf{W}),
\]
%
which is the optimal recovery of $\mathbf{A} \oplus \mathbf{D}$. The
authentication decision is made via the following threshold test,
%
\[
\hat{\theta} = d(\hat{\mathbf{W}}) \mathop{\lessgtr}^1_0 \tau n,
\]
%
which accepts or rejects based on the closeness of the probe biometric
feature vector to the enrollment biometric feature vector.
\end{comment}

\begin{comment}
Since the channel codes are operating under capacity, we will assume
that they have a positive error exponent $E(R) > 0$ and that the
probability of decoding error when using these codes on a BSC with
crossover probability $p$ is bounded by
%
\[
P_e \leq 2^{-n E(R) + o(n)}.
\]
%
It is well known that there exist code constructions that support
these assumptions \cite{Gallager-1968}. The parity check matrix
$\mathbf{H}$ is fixed and full row rank. Hence $\mathbf{H}\mathbf{A}$
is i.i.d.\ Bernoulli($0.5$).
\end{comment}

\subsection{Equivalence of Fuzzy Commitment and Secure Sketch}
\label{sec:accuracyanalysis}

\begin{comment}
In this section, we analyze the FRR, FAR, and SAR of all of our
systems.  We will first analyze the FRR $P_{FR}$ and the FAR $P_{FA}$
for the keyless secure sketch system.  We will then argue that these
quantities are the same for the two-factor secure sketch system and
both variants of the fuzzy commitment system.  Then, for each system
and the various scenarios of data exposure, we will show that the SAR
$P_{SA}$ is equal to either $P_{FA}$ or one.
\end{comment}

We now develop an equivalence between the properties of the fuzzy
commitment and secure sketch schemes presented in the previous two
subsections. We show the conceptual equivalence between the two
architectures and derive expressions for the performance metrics
defined in Section~\ref{subsec:metrics}, showing that the performance is
the same.

Reviewing the decoding procedures of fuzzy commitment and secure
sketch one sees that the procedures are nearly identical.  The
authentication decision is determined by whether or not
$\hat{\mathbf{W}}$, the lowest Hamming weight sequence in a given
coset, has Hamming weight greater or less than $\tau n$.  The coset is
specified by its syndrome and the only difference between the systems
is how this syndrome is computed.

In the two-factor secure sketch system, the syndrome is specified as
\begin{align}
q_{_{SS}}(\mathbf{D}, \mathbf{L}, \mathbf{S}) &= \mathbf{H}\mathbf{D}
\oplus \mathbf{L} \oplus \mathbf{S} \notag \\
 &= \mathbf{H}(\mathbf{A} \oplus
\mathbf{D}) \oplus \mathbf{K} \oplus \mathbf{L} \label{eq:decodingsyndromeSS}
\end{align}
In the two-factor fuzzy commitment system, the syndrome is specified
as
\begin{align}
q_{_{FC}}(\mathbf{D}, \mathbf{L}, \mathbf{S}) &= \mathbf{H}(\mathbf{D}
\oplus \mathbf{L} \oplus \mathbf{S}) \notag \\ 
&= \mathbf{H}(\mathbf{A}
\oplus \mathbf{D}) \oplus \mathbf{H}\mathbf{G}^\mathrm{T}\mathbf{Z}
\oplus \mathbf{H}(\mathbf{K} \oplus \mathbf{L}) \notag\\ 
&=\mathbf{H}(\mathbf{A} \oplus \mathbf{D}) \oplus \mathbf{H}(\mathbf{K}
\oplus \mathbf{L})  \label{eq:decodingsyndromeFC}
\end{align}
The decision for $\hat{\theta}$ is a deterministic function of the
syndrome, defined identically for both systems.

%The syndrome is only a function of $\mathbf{H}(\mathbf{A} \oplus
%\mathbf{D})$ and $(\mathbf{K},\mathbf{L})$.

\begin{comment}
In the keyless secure sketch system, the syndrome is
%
\begin{align*}
q(\mathbf{S}, \mathbf{D}) &= \mathbf{H}\mathbf{D} \oplus \mathbf{S} \\
&= \mathbf{H}(\mathbf{A} \oplus \mathbf{D}).
\end{align*}
%
In the keyless fuzzy commitment system, the syndrome is
%
\begin{align*}
q(\mathbf{S}, \mathbf{D}) &= \mathbf{H}(\mathbf{D} \oplus \mathbf{S})
\\ &= \mathbf{H}(\mathbf{A} \oplus \mathbf{D}) \oplus
\mathbf{H}\mathbf{G}^\mathrm{T}\mathbf{Z} \\ &= \mathbf{H}(\mathbf{A}
\oplus \mathbf{D}).
\end{align*}
%
\end{comment}

In both systems, during the authentication of the legitimate user,
where $\mathbf{D} = \mathbf{B}$ and $\mathbf{L} = \mathbf{K}$, the
computed syndrome is identical and equal to $\mathbf{H}(\mathbf{A}
\oplus \mathbf{B})$.  Note that this is true of both keyed and keyless
variants of the systems.  Thus, the distribution of $\hat{\theta}$
given $\theta = 1$ is identical for both types of systems and thus the
FRR is identical.

In determining the FAR -- the case of an attack by an uninformed adversary -- the
input vectors $(\mathbf{D},\mathbf{L}) = (\mathbf{C},\mathbf{J})$ can
have an arbitrary joint distribution, but must be independent of the pair
$(\mathbf{A},\mathbf{K})$.  Regardless of the distribution of
$(\mathbf{C},\mathbf{J})$, the syndrome in both systems is i.i.d.\
Bernoulli($0.5$), since $\mathbf{A}$ is assumed to be an independent
i.i.d.\ Bernoulli($0.5$) sequence and $\mathbf{H}$ has full row rank
(cf. Lemma~\ref{lemm:linInd} below). Since the syndromes are equal in 
distribution for both systems, the
authentication decisions $\hat{\theta}$ are also equal in distribution
for both systems, and hence the FAR performance is the same.

Determining the SAR of these systems requires consideration of
scenarios when the adversary has access to $\mathbf{A}$, $\mathbf{S}$,
and/or $\mathbf{K}$.  In contrast to the scenario considered for the
FAR analysis, the availability of this additional information may
allow the adversary to alter the distribution of the decoding
syndrome.  However, as we will see in
Theorem~\ref{thm:AccuracyAnalysis} below, the SAR for
secure sketch and fuzzy commitment is also the same.

Before we proceed, consider the following result that will be
useful in understanding and proving some of the theorems
that follow:

\begin{lem} \label{lemm:linInd}
Let $\bfA$ be a length-$n$ i.i.d. Bernoulli-$(0.5)$ random vector and
let $\bfH$ and $\tilde{\bfH}$ be, respectively, $m \times n$ and
$\tilde{m} \times n$ full row-rank binary matrices whose rows are
linearly independent of each other.  Then, for any pair of binary
vectors, $\bfs$ and $\tilde{\bfs}$, of lengths $m$ and $\tilde{m}$
respectively, $\Pr[\bfH \bfA = \bfs | \tilde{\bfH} \bfA =
  \tilde{\bfs}] = \Pr[\bfH \bfA = \bfs ] = 2^{-m}$.
\end{lem}

The proof of this lemma appears in Appendix~\ref{app:lemm:linInd}.
Note that, since the channel codes are assumed to operate at a rate $R
= k/n$ which is \emph{below} capacity they have a positive {\em error
  exponent} $E(R) > 0$.  This means that the probability of decoding
error when using these codes on a BSC-$p$ is bounded as
\[
P_e \leq 2^{-n E(R) + o(n)}.
\]
where $E(R) = \min_q\,( D(q\|p) + \max\{1 - h_b(q) - R, 0\})$ and the KL
divergence between two Bernoulli distributions, Bernoulli($q$) and
Bernoulli($p$) is defined as
\[
D(q\|p) := q \log_2 \frac{q}{p} + (1-q) \log_2 \frac{1-q}{1-p}.
\]
It is well known that, for sufficiently large $n$, there exist code constructions that satisfy
these properties \cite{Gallager-1968}.

\begin{thm} \label{thm:AccuracyAnalysis}
The FRR and FAR of both keyed and keyless variants of fuzzy commitment
and secure sketch is the same and is bounded as
\begin{enumerate}
\item[(i)] $P_{FR} \leq 2^{-n D(\tau\|p)} + 2^{-n E(R) + o(n)}$, 
\item[(ii)] $P_{FA} \leq 2^{-n \left( \frac{m}{n} - h_b(\tau) \right)
}$.
\end{enumerate}
The SAR of the two-factor (keyed) fuzzy commitment and secure sketch
schemes for various cases of data exposure are identical and given by
\begin{enumerate}
\item[(iii)] $P_{SA}(\mathbf{K})  = P_{FA}$,
\item[(iv)] $P_{SA}(\mathbf{A}) = P_{FA}$,
\item[(v)] $P_{SA}(\mathbf{S}) = P_{SA}(\mathbf{A},\mathbf{K}) =
  P_{SA}(\mathbf{A},\mathbf{S}) = P_{SA}(\mathbf{S},\mathbf{K}) =
  P_{SA}(\mathbf{A},\mathbf{S},\mathbf{K})=1$.
\end{enumerate}
The SAR of the keyless fuzzy commitment and secure sketch schemes for
various cases of data exposure are identical and given by
\begin{enumerate}
\item[(vi)] $P_{SA}(\mathbf{S}) = P_{SA}(\mathbf{A}) =
P_{SA}(\mathbf{A},\mathbf{S}) = 1$.
\end{enumerate}
\end{thm}
Please refer to Appendix~\ref{app:thm:AcuracyAnalysis} for the proof
of the theorem. In parts (i) and (ii) the theorem characterizes
exponentially decaying upper bounds on the FRR and FAR, and hence also
lower bounds on the exponents.  In order to obtain these exponentially
decaying bounds, the operating parameters must satisfy the previously
listed assumptions, that is, $0.5 > \tau > p$ and $m/n > h_b(\tau)$.
Note that for all of our systems, knowledge of the stored data
$\mathbf{S}$ drastically improves the ability of the adversary to gain
access. For all of our systems, the SAR is equal to one for an
adversary enhanced with the knowledge of $\mathbf{S}$, cf. parts~(v)
and~(vi) above.  This is because, as is formalized in the proof, an adversary
with knowledge of $\mathbf{S}$ can gain access by choosing
$(\mathbf{C},\mathbf{J})$ based on knowledge of $\mathbf{S}$ so that
the decoding coset contains a low-weight error sequence with
probability one.  In fact, this limitation is not unique to ECC-based
systems as the following theorem shows.

\begin{thm} \label{thm:SARgivenS}
For any two-factor system,
\begin{enumerate}
\item[(i)] $P_{SA}(\mathbf{S}) \geq 1 - P_{FR}$.
\end{enumerate}
If for every $\mathbf{S} \in \mathcal{S}$, there exist
$\mathbf{D},\mathbf{L}$ such that $g(\mathbf{D},\mathbf{L},\mathbf{S})
= 1$, then
\begin{enumerate}
\item[(ii)] $P_{SA}(\mathbf{S}) = 1$.
\end{enumerate}
\end{thm}
The proof appears in Appendix~\ref{app:thm:SARgivenS}.

Fuzzy commitment and secure sketch also have identical privacy leakage
as demonstrated by the following theorem.
\begin{thm} \label{thm:InfoLeakage}
In the two-factor fuzzy commitment and secure sketch systems, the
privacy leakage of $\mathbf{A}$ from $\mathbf{S}$, from $\mathbf{K}$,
or from $(\mathbf{S},\mathbf{K})$ is, respectively,
\begin{enumerate}
\item[(i)] $I(\mathbf{A};\mathbf{K}) = 0$,
\item[(ii)] $I(\mathbf{A};\mathbf{S}) = 0$,
\item[(iii)] $I(\mathbf{A};\mathbf{S},\mathbf{K}) = m = n(1 - R) > 0$.
\end{enumerate}
In the keyless variant of fuzzy commitment and secure sketch the
privacy leakage of $\mathbf{A}$ from $\mathbf{S}$ is
\begin{enumerate}
\item[(iv)] $I(\mathbf{A};\mathbf{S}) = m = n(1 - R) > 0$.
\end{enumerate}
\end{thm}
The proof of this theorem is given in
Appendix~\ref{app:thm:InfoLeakage}. From an authentication
perspective, it is interesting that the additional independent source
of randomness ${\bf Z}$ in fuzzy commitment based systems does not
improve the privacy leakage properties in comparison to secure sketch
based systems where such randomness is unavailable.

The fuzzy commitment and secure sketch systems are equivalent in terms
of many performance metrics but they differ in terms of storage and
key length requirements.  The fuzzy commitment system requires $n$
bits to store the data since $H(\mathbf{S}) = n$.  It also uses an
$n$-bit key to mask the stored data in the two-factor variant.  On the
other hand, secure sketch system requires only $m$ bits for storage
since $H(\mathbf{S}) = m$ due to the fact that only the syndrome of
$\mathbf{A}$ is being stored.  Similarly, it also uses only an $m$-bit
key to mask the stored data in the two-factor variant.

\begin{table*}
\center
\renewcommand{\arraystretch}{1.5}
\begin{tabular} {| r || c | c | c | c |}
\hline
\multirow{2}{*}{\bf System} & \multicolumn{2}{|c|}{\bf Keyless} &
\multicolumn{2}{|c|}{\bf Two-factor} \\
\cline{2-5} & {\bf Fuzzy Commitment} & {\bf Secure Sketch} & {\bf
 Fuzzy Commitment} & {\bf Secure Sketch} \\
\hline \hline {\bf False Rejection Rate} &
\multicolumn{4}{|c|}{$P_{FR} \leq 2^{-n D(\tau\|p)} + 2^{-n E(R) +
o(n)}$} \\
\hline {\bf False Acceptance Rate} & \multicolumn{4}{|c|}{$P_{FA} \leq
2^{-n \left( \frac{m}{n} - h_b(\tau) \right) }$} \\
\hline \multirow{2}{*}{\bf Successful Attack Rate} &
\multicolumn{2}{|c|}{\multirow{2}{*}{$P_{SA}(\mathbf{A}) =
P_{SA}(\mathbf{S}) = 1$}} & \multicolumn{2}{|c|}{$P_{SA}(\mathbf{A}) =
P_{SA}(\mathbf{K}) = P_{FA}$} \\
 & \multicolumn{2}{|c|}{} & \multicolumn{2}{|c|}{$P_{SA}(\mathbf{S}) =
 P_{SA}(\mathbf{A},\mathbf{K}) = 1 $} \\
\hline \multirow{2}{*}{\bf Privacy Leakage} &
\multicolumn{2}{|c|}{\multirow{2}{*}{$I(\mathbf{A};\mathbf{S})=m$}} & \multicolumn{2}{|c|}{$I(\mathbf{A};\mathbf{K}) = 0$,
$I(\mathbf{A};\mathbf{S},\mathbf{K}) = m$} \\
 & \multicolumn{2}{|c|}{} & \multicolumn{2}{|c|}{$I(\mathbf{A};\mathbf{S}) = 0$} \\
\hline {\bf Storage Requirements} & $H(\mathbf{S}) = n$ &
$H(\mathbf{S}) = m$ & $H(\mathbf{S}) = H(\mathbf{K}) = n$ &
$H(\mathbf{S}) = H(\mathbf{K}) = m$ \\
\hline
\end{tabular}
\caption{Summary and Comparison of System Performance}
% \caption{Summary and comparison of system performance for keyless and two-factor variants of fuzzy commitment and secure sketch schemes.}
\label{tab:SystemComp}
\end{table*}
\section{Linkage Resistance and Revocability Properties}
\label{sec:revocability}

In this section we consider two desirable properties for secure
biometrics -- revocability and resistance to linkage attacks -- and
study them in the context of noisy enrollments at multiple access
control devices. We will only consider two-factor systems in this
section. Although the results to be presented in this section apply
equally to both secure sketch and fuzzy commitment based systems,
proofs will be provided only for secure-sketch based systems since the
two types of systems are performance-equivalent as discussed in
Section~\ref{sec:accuracyanalysis}.

Revocability is the ability to tolerate partial compromises of data.
By~\emph{partial} compromise we mean that, in a two-factor access
control system, either the key or the stored data has been revealed to
the adversary, but {\em not} both. On the other hand, we say that a
two-factor system is~\emph{fully} compromised if both the key and the
stored data have been revealed to the adversary.  A secure biometric
is said to be revocable if, given knowledge of a partial compromise,
the user or a system administrator can delete certain data and
establish a new enrollment based on the same biometric without any
loss in privacy or authentication performance.

Linkage attacks can occur in situations where the same biometric is
used to enroll in multiple biometric systems, e.g., on several access
control devices.  If an adversary compromises a subset of the devices,
the compromised data can be used to attack the remaining devices. The
compromised data can both leak information about the underlying
biometric and can be exploited to mount a successful attack, i.e.,
gain unauthorized access to, one of the remaining devices.

\subsection{Performance Measures for Multiple Biometric Systems}
\label{sec:MultSys}

We now present our model for parallel enrollment across multiple
biometric systems.  We assume that the biometric in question has been
enrolled in $u$ systems. Each of the $u$ biometric systems has an
enrollment vector.  These vectors, $\bfA_i$, $i \in \{1, \ldots,u\}$,
are related in a conditionally independent manner to a common
underlying biometric $\bfA_0$ according to the measurement model in
Section~\ref{sec:framework}. In other words,
\[
P_{\bfA_i|\bfA_0}(\bfa_i|\bfa) = (1-p_i)^{n - d_H(\bfa_i,\bfa)}
p_i^{d_H(\bfa_i,\bfa)}
\]
where $p_i \in [0,0.5)$, all vectors are binary and $d_H(\cdot,
\cdot)$ is the Hamming distance between its arguments. For
convenience, we define $p_0=0$.  Encoding and decoding functions
$\{F_i(\cdot), g_i(\cdot)\}_{i=1}^u$ are paired and need not be
identical for all systems. At enrollment, each system $i \in \{1,
\ldots,u\}$ observes $\bfA_i$, and the stored data and key for system
$i$ are generated as $(\bfS_i, \bfK_i) = F_i(\bfA_i)$.  The joint
distribution across the $u$ systems is given by
\begin{equation}
P_{\bfS^u, \bfK^u,\bfA_0} (\bfs^u,\bfk^u,\bfa) = P_{\bfA_0}(\bfa)\,\prod_{i=1}^u
P_{\bfS_i,\bfK_i|\bfA_0} (\bfs_i,\bfk_i|\bfa)
, \label{eq.multEnrollModel}
\end{equation}
where
\[
P_{\bfS_i,\bfK_i|\bfA_0} (\bfs_i,\bfk_i|\bfa) = \sum_{\mathbf{a}_i}  \Pr \big[ F_i(\bfa_i)
  = (\bfs_i,\bfk_i) \big] P_{\bfA_i|\bfA_0}(\bfa_i|\bfa),
\]
and $\bfS^u$ and $\bfK^u$ are respectively the $u$-tuples of stored
data and key vectors.

%Analyzing robustness to linkage attacks and revocability properties
%requires the consideration of $u$ parallel enrollments with respect to
%the same underlying biometric.  The multiple enrollments can be viewed
%as belong to either parallel authentication services that are based on
%the same biometric or a series of re-enrollments for a single
%authentication service following revocations due to compromises of
%previous enrollments.

Recall from the discussion of Sec.~\ref{sec:framework}
(cf.~Fig.~\ref{fig:framework}) that the legitimate user of system $i$
will try to authenticate using $(\bfB, \bfK_i)$ while an adversary
will use some $(\bfC, \bfJ)$. The crossover probability of
system-$j$'s probe channel will be denoted by $\alpha_j \in
[0,0.5)$. The FRR and FAR are, respectively, given by
\begin{eqnarray*}
P_{FR}(i) &:=& \Pr \big[ g_i(\bfB,\bfK_i,\bfS_i) = 0 \big], \\ 
P_{FA}(i) &:=& \max_{p_{\bfC,\bfJ}} \Pr \big[ g_i(\bfC,\bfJ,\bfS_i) = 1 \big],
\end{eqnarray*}
which are the same as the definitions for a single system in isolation.

In contrast, the existence of multiple systems necessitates the
generalization of the definition of SAR, in order to account for
compromises across multiple biometric systems. Expanding upon the
framework of Sec.~\ref{sec:framework}, we define ${\cal V}$ to be a
subset of $\{\bfS_1, \bfK_1, \bfS_2, \bfK_2, \ldots, \bfS_u,
\bfK_u\}$.  Equivalently we write ${\cal V} = \cup_{i=1}^u {\cal V}_i$
where ${\cal V}_i \subseteq \{\bfS_i, \bfK_i\}$, possibly the empty
set.  Also, to be able to study the effect of compromised enrollment
biometrics, we define the set ${\cal A}$ to be a subset of $\{\bfA_0,
\bfA_1, \bfA_2, \ldots, \bfA_u\}$.

Given knowledge of ${\cal V}$ and ${\cal A}$ by an adversary, the SAR
against system $i$ is
\[
P_{SA}(i,{\cal V},{\cal A}) = \max_{p_{\bfC,\bfJ|{\cal V}, {\cal A}}} \Pr \big[
  g_i(\bfC,\bfJ,\bfS_i) = 1 \big].
\]
%
%When multiple parallel enrollments are compromised, the natural
%extension for measuring the information leakage is
%$I(\bfA_0;\mathbf{V})$, where $\mathbf{V}$ is a subset of
%$\{(\bfA_1,\bfS_1, \bfK_1), \ldots,
%(\bfA_u,\bfS_u, \bfK_u)\}$.  

%\begin{figure}
%\centering
%\includegraphics[width=2.5in,height=2.7in]{figures/biometrics-ceostyle-attacks.pdf}
%\caption{An attacker may corrupt multiple devices, where compromising
%  a device entails gaining access to the stored data, or the secret
%  key or both.}
%\label{fig:bioceo-attacks}
%\end{figure}

\subsection{Privacy Leakage Across Multiple Systems}
\label{sec:LeakageMultiple}

In this section we give a tight characterization of the privacy
leakage, i.e., the amount of information leaked about the user's
biometric when some subset of the stored data is compromised.  In the
analysis that follows, we assume that all $u$ biometric systems are
secure sketch-based systems with parity check matrices $\bfH_1,
\ldots, \bfH_u$ which may have different row-sizes but the same
column-size. As we have already proved the equivalence between secure
sketch and fuzzy commitment in Section~\ref{sec:Systems}, the results
derived for multiple secure sketch-based systems immediately extend to
multiple fuzzy commitment-based biometric systems.  In other words,
statements about the parity check matrices $\bfH_i$ can be
appropriately modified into similar statements about the generator
matrices ${\bf{G}}_i$ used in fuzzy commitment-based systems.

While deriving the privacy leakage, we also state simplifications for
a number of interesting special cases.  In particular we consider both
the ``noiseless'' enrollment case where $\bfA_0 = \bfA_1 = \ldots =
\bfA_u$ and the ``identical'' enrollment function case where all
systems use the same ECC, i.e., $\bfH_1 = \ldots = \bfH_u$.  We also
write $\mathrm{rank}(\bfH_1,\ldots,\bfH_j)$ to denote the rank of
$[\bfH_1^T,\ldots,\bfH_j^T]$.

Our main result connects the amount of information leakage with an
easily-characterized rank property of the parity check matrices of the
compromised systems.
\begin{thm} \label{thm:MultInfoLeak} 
Given the enrollment model of~(\ref{eq.multEnrollModel}), assume,
without loss of generality, an ordering of the systems such that for
some index $l$, $0 \leq l \leq u$, ${\cal V}_i = \{\bfS_i, \bfK_i\}$
for all $i \in [1,l]$ and ${\cal V}_i \subset \{\bfS_i, \bfK_i\}$ for
all $i > l$.  Then, the information about $\bfA_0$ leaked by ${\cal V}
= \cup_{i=1}^u {\cal V}_i$ is
\begin{equation*}
I(\bfA_0 ; {\cal V}) = \left\{
\begin{array}{ccc}
0 & \mbox{if} &  l = 0\\ I(\bfA_0 ;
\bfH_1\bfA_1,\ldots,\bfH_l\bfA_l) & \mbox{else}
\end{array} \right..
\end{equation*}
Additionally:

\begin{enumerate}
\item[(i)] In general,
\begin{equation*}
I(\bfA_0 ; \cV) \leq \mathrm{rank}(\bfH_1,\ldots,\bfH_l).
\end{equation*}
\item[(ii)] For noiseless, non-identical enrollment functions,
\begin{align*}
I(\bfA_0 ; \cV) &= \mathrm{rank}(\bfH_1,\ldots,\bfH_l).
\end{align*}
while for the identical enrollment function case with $l \geq 1$, we have 
\[
I(\bfA_0 ; \cV) = \mathrm{rank}(\bfH_1).
\]
\end{enumerate}
\end{thm}

The proof of this theorem is given in
Appendix~\ref{app:thm:MultInfoLeak}.  Importantly, this theorem tells
us that information about the underlying biometric is leaked only if
there is at least one fully compromised system (i.e., $l > 0$). Hence,
unless both the key and stored data of a particular system have been
compromised, that system can be revoked by erasing the un-compromised
data (e.g., the key if the stored data has been leaked).  The theorem
indicates that biometric measurement noise can only help mask the
private data.  To see this, consider the case $l>0$ and note that if
the enrollment noise is high enough, the information between $\bfA_0$
and $\bfH_1 \bfA_1, \ldots \bfH_l \bfA_l$ can be quite small,
certainly smaller than when there is no enrollment noise. This last
statement follows from the information processing inequality which
tells us that the privacy leakage when enrollments are noisy is upper
bounded by the privacy leakage when enrollments are noiseless.
% In this situation the relation in
% statement-(i) is met with equality.

Part (i) also tells us that the privacy leakage depends on the rank of
the matrix formed by stacking the parity check matrices of the fully
compromised systems.  We term this the ``collective'' rank of the set
in question.  The collective rank is at most equal to the sum of the
ranks of the individual parity check matrices and will be strictly
less if there is linear dependence between the rows of the matrices.
Further, as part (ii) tells us, in the special case of noiseless
enrollments we can make an exact statement about privacy leakage in
terms of collective rank.  Finally, in the special case of noiseless
enrollments and identical enrollment functions, the first fully
compromised system leaks all the information there is to be leaked
about the underlying biometric.

We can sketch a candidate design rule arising from these results.  To
obtain a set of systems that minimize the privacy leakage in the face
of the compromise of some subset of the stored data and keys, the
collection of parity check matrices should be designed to minimize the
linear dependencies across the matrices.  Of course, at the same time
the matrices must individually specify good error correcting codes,
else the false rejection rate would be too high.  However, such
minimal privacy leakage comes at a cost.  Further, to achieve minimum
collective rank, one should simply use the same parity check matrix
for each system.  However, as we discuss in the next subsection, this
choice makes the remaining uncompromised systems more vulnerable to false
authentications.  Thus, if we design the multiple systems to minimize
privacy leakage, we pay a price in terms of the security of the
individual systems.

\subsection{Authentication Attacks with Multiple Systems}
\label{sec:SARMultiple}

In situations where some subset of systems based on the same biometric
have been compromised, an attacker may be able to use the compromised
data to enhance his ability to authenticate falsely.  The following
theorem states results on the successful attack rates for our
two-factor secure biometric systems.  The theorem is proved in
Appendix~\ref{app:thm:MultiSAR1}.

\begin{thm} \label{thm:MultiSAR1}
Let $u$ noisy, non-identical enrollments be generated for a secure
two-factor biometric system (fuzzy commitment or secure sketch).
Consider any system $j \in \{1,\ldots,u\}$.
\begin{enumerate}
\item[(i)] If either $\bfS_j \in \cV_j$ or both $\bfA_j \in {\cal
  A}$ and $\bfK_j \in \cV_j$, then
\[
P_{SA}(j,\cV,{\cal A} ) = 1.
\]
%\item[(ii)] If both $\bfA_0 \in {\cal A}$ and $\bfK_j \in \cV_j$, then
%\[
%P_{SA}(j,\cV,{\cal A} ) \geq 1 - P_{FR}(j).
%\]
%
\item[(ii)] If $\bfK_j \in \cV_j$ and for some $i \neq j$, $\bfA_i \in
  {\cal A}$ and $p_i \leq p_j$ then
\[
P_{SA}(j,\cV,{\cal A} ) \geq 1 - P_{FR}(j).
\]
\item[(iii)] If $\cV_j = \{ \}$, the null set, then
\[
P_{SA}(j,\cV,{\cal A}) = P_{FA}(j).
\]
\item[(iv)] If $\bfS_j \notin \cV_j$, ${\cal A} = \{ \}$, and $\cV_i
  \subset \{\bfS_i, \bfK_i\}$ for each $i \neq j$ then
\[
P_{SA}(j,\cV,{\cal A} ) = P_{FA}(j).
\]
\end{enumerate}
\end{thm}

In part (i), an adversary who has access to the stored data of the
target system can easily find a low-weight element of the coset
corresponding to $\bfS_j$, yielding access with probability one as per
Theorem~\ref{thm:AccuracyAnalysis}(v).\footnote{Note that in this part
  if only $\bfS_j$ is leaked, but not $\bfK_j$, then this is a
  revocable scenario, i.e., the old $\bfK_j$ can be revoked and a new
  key assigned. Until this is done, however, the probability of
  successful attack is one, as given above. Once $\bfK_j$ is revoked,
  the probability of successful attack becomes $P_{FA}(j)$.}

In part (ii), the adversary has access to the key of the system to be
attacked and at least one enrollment biometric of some other system
$\bfA_i$ or the ground truth biometric $\bfA_0$.  In these settings we
show that the adversary can use this data to imitate a probe biometric
of the legitimate user and launch an authentication attack with a high
probability of success.

In contrast, in parts (iii)--(iv), the adversary cannot do better than
the nominal false acceptance rate.  In part (iii), neither the key nor the
stored data of the target system are compromised, but the $\cV_i$ for
$i \neq j$ can be arbitrary.  Then, because $\bfK_j$ is independent of
all other parts of the system, the attacker cannot improve his
probability of success over that of random guessing.  In part (iv),
the key of the target system may be compromised, but in all other
systems only a strict subset of the data is compromised (either just
the stored data, just the key, or neither) and, further, no enrollment
biometrics are compromised.  In this situation the adversary is again
not able to authenticate with probability higher than the FAR.

The following theorem considers the effect of the joint structure of
the parity check matrices employed on different access control devices on
the probability of successful attack.  It establishes that if certain
joint structure is present, the adversary can leverage this structure
to improve dramatically the likelihood of being able to falsely
authenticate on uncompromised systems.  The theorem is proved in
Appendix~\ref{app:thm:MultiSAR2}.

\begin{thm} \label{thm:MultiSAR2}
Given the enrollment model of~(\ref{eq.multEnrollModel}), assume the
two-factor systems are ordered such that there is some index $l$, $1
\leq l \leq u$ such that ${\cal V}_i = \{\bfS_i, \bfK_i\}$ for all $i
\leq l$ and ${\cal V}_i = \bfK_i$ for all $i > l$. Let ${\cal A} = \{
\}$. Now, consider any system index $j \geq l+1$.
\begin{enumerate}
\item[(i)] For noiseless enrollments if $\mathrm{rank}(\bfH_1, \ldots,
  \bfH_l, \bfH_j) = \mathrm{rank}(\bfH_1, \ldots, \bfH_l)$ then
\[
P_{SA}(j,\cV,{\cal A}) = 1.
\]
\item[(ii)] For noisy enrollments if $\mathrm{rank}(\bfH_1, \ldots,
  \bfH_l, \bfH_j) = \mathrm{rank}(\bfH_1, \ldots, \bfH_l)$ and for all
  $0 \leq i \leq u$, $p_i \leq \alpha_j$, where $\alpha_j$ is the
  crossover probability of system-$j$'s probe channel, then
\[
P_{SA}(j,\cV,{\cal A}) \geq 1 - P_{FR}(j).
\]
\item[(iii)] If $\mathrm{rank}(\bfH_1, \ldots, \bfH_l, \bfH_j) =
  \mathrm{rank}(\bfH_1, \ldots, \bfH_l) + \mathrm{rank}(\bfH_j)$ then
  (in either the noisy or noiseless case)
\[
P_{SA}(j,\cV,{\cal A}) = P_{FA}(j).
\]
\end{enumerate}
\end{thm}

The conditions in the first two parts of Thm.~\ref{thm:MultiSAR2} mean
that the row space of $\bfH_j$ lies within the span of the rows of
$\bfH_1,\ldots,\bfH_l$.  In this situation, an attacker can gain access
with high probability.  In contrast, if the parity check matrix
$\bfH_j$ used to define the stored data in the system under attack is
linearly independent of the matrices defining the compromised systems,
then the compromised data is useless in attempts to improve the
successful attack rate beyond the nominal false acceptance rate of the
system.

%{\bf scd: need to add discussion around following example connecting
 % them better to theorems}

To build intuition, we study the implications of
Thm.~\ref{thm:MultiSAR2} through a sequence of examples.  In keeping
with our previous development, we consider secure sketch-based
biometric systems, though the results translate to fuzzy
commitment-based access control devices as well. In each example we
consider three biometric systems, $u = 3$.  The three enrollment
matrices $\bfH_1$, $\bfH_2$, $\bfH_3$, are each of size $m \times n$
and full rank $m$ where $n = 3m$.  We consider an adversary that is
trying to authenticate with respect to system \#3, having gained
access to all data {\em except} $\bfS_3$, i.e., $\cV = \{\bfS_1,
\bfK_1, \bfS_2, \bfK_2, \bfK_3\}$.  In some of the examples, we will
find it useful to refer back to Lemma~\ref{lemm:linInd}, which relates
linear independence between the rows of the parity check matrices to
statistical independence of the syndromes $\bfH_i \bfA_i$.

\begin{example} \label{ex.A} (noiseless enrollments) {\em
Consider noiseless enrollments, $\bfA_0 = \bfA_1 = \bfA_2 = \bfA_3$
and $\bfH_3 = \bfH_1 \oplus \bfH_2$.  In this setting, using the
elements of $\cV$, the adversary can calculate the stored data 
%decrypted syndrome
of the third system as $\bfS_3 = \bfS_1 \oplus \bfS_2 \oplus \bfK_1
\oplus \bfK_2 \oplus \bfK_3$.  Picking $\bfC$ ($= \bfD$) such that
$\bfH_3 \bfC = \bfS_3$ and setting ${\bf J} = {\bf L} = {\bf 0}$ the
all-zeros syndrome, the adversary can force the decoder to the coset
containing the all-zeros vector.  Recall that the decoder looks for
the lowest weight vector in the set $\bfH_3 \bfD \oplus \bfS_3 \oplus
{\bf L}$.  The probability of success of this attack is one.}
\end{example}

\begin{example}\label{ex.B} (identical enrollment functions) {\em 
Consider the setting where $\bfH_1 = \bfH_2 = \bfH_3$.  If enrollments
are noiseless then, e.g., $\bfS_3 = \bfS_1 \oplus \bfK_1 \oplus
\bfK_3$ and the attack of Example~\ref{ex.A} works, allowing the
adversary to successfully access system \#3 with probability one. In
fact compromising the stored data and key of any single system will
allow an attacker to access any other system whose key is compromised
with probability one. If enrollments are noisy but $p_1 = p_2 = p_3$
then $\bfS_1 \oplus \bfK_1$ specifies a coset that contains a vector
close to $\bfA_0$.  Pick {\em any} element of this coset as $\bfD$ and
use $\bfK_3$ for ${\bf L}$.  These choices will yield the same
probability of successful attack as a legitimate probe generated from
$\bfA_0$, i.e., at least $1 - P_{FR}$.}
\end{example}

\begin{example} \label{ex.C} (linearly independent enrollment functions) {\em 
Now consider the case when the rows of $\bfH_1$, the rows of $\bfH_2$,
and the rows of $\bfH_3$ are all linearly independent of one another.
Then, by Lemma~\ref{lemm:linInd}, whether or not enrollments are
noisy, the information about the biometric leaked by the compromised
data is independent from $\bfS_3$.  Hence, the compromised data does
not enhance the adversary's ability to authenticate falsely.}
\end{example}

Example~\ref{ex.C} suggests that a cross-system design of the codes,
i.e., $\bfH_1, \ldots, \bfH_u$, that minimizes the linear dependence
between parity check matrices can obviate the danger of linkage
attacks.  However, it is not always possible to design fully
independent parity check matrices while maintaining the desired full
rank of each.  This is due to dimensionality restrictions.  In the
examples, $m = n/3$.  Thus, if we added another biometric system,
i.e., $u=4$, maintaining full linear independence is not possible.

\begin{example} \label{ex.D} (partially linearly dependent enrollment functions) {\em
Theorem~\ref{thm:MultiSAR2} considers the two extreme cases of linear
dependence between the parity check matrix $\bfH_j$ of the system
under attack and those of the compromised systems, $\bfH_1, \ldots,
\bfH_l$.  Full linear dependence is considered in parts (i) and (ii)
of the theorem, and full linear independence in part (iii).  In this
example we consider an intermediate scenario of {\em partial} linear
dependence.

In particular, let $\bfH_a$, $\bfH_b$, $\bfH_c$, $\bfH_d$ be full-rank
$m/2 \times n$ matrices where all of the rows are linearly
independent.  Let $\bfH_1^T = [\bfH_a^T \, \bfH_b^T]$, $\bfH_2^T =
[\bfH_a^T \, \bfH_c^T]$, and $\bfH_3^T = [\bfH_a^T \, \bfH_d^T]$.
Again let $\cV = \{\bfS_1, \bfK_1, \bfS_2, \bfK_2, \bfK_3\}$.  The first
half of the vector $\bfS_3 \oplus \bfK_3$ equals $\bfH_a \bfA$, which,
for noiseless enrollments, is the same as the first half of the
$\bfS_1 \oplus \bfK_1$ and $\bfS_2 \oplus \bfK_2$ vectors, both of
which can be calculated from the stored information.  However, by
Lemma~\ref{lemm:linInd} the second half of the $\bfS_3 \oplus \bfK_3$
vector is statistically independent of all compromised data.

We now describe a natural attack on the system descried in
Example~\ref{ex.D}.  First note that, in the same manner as in the
earlier examples, the attacker can set the first half of the syndrome
arbitrarily.  One attack would be to pick these $m/2$ constraints to
eliminate as {\em few} low-weight sequences as possible.  Ideally,
these constraints would be picked so that, regardless of the remaining
$m/2$ bits of the syndrome, each possible coset (after all $m$
syndrome bits are set) would contain at least one low-weight sequence
(i.e., a sequence with fewer than $\tau n$ ones).  Whether such an
attack is possible depends on the specific $\bfH_a$ and $\bfH_d$
matrices.  One should note that low-weight sequences are not uniformly
distributed over the cosets of $\bfH_a$.  This means that even
determining whether such an attack is possible for specific $\bfH_a$
and $\bfH_d$ matrices is likely quite computationally challenging.
These considerations illustrate the difficulty of determining the SAR
in these settings.  At a minimum, we can say that the SAR must be at
least as large as the FAR.  This follows since the attacker can make
the SAR equal to the FAR simply by ignoring the compromised data and
setting all $m$ syndrome bits at random.

}
\end{example}

\subsection{Formulation of ECC Design Problem for Multiple Systems}
\label{sec:MultiDesign}

In the previous two subsections, we analyzed information leakage and
authentication attacks when an adversary has compromised multiple
enrollments based on the same underlying biometric.
Theorems~\ref{thm:MultInfoLeak} and~\ref{thm:MultiSAR1} tell
us that unless there are fully compromised systems, no information is
leaked about the underlying biometric and there is no way to improve
the probability of successful attack beyond the nominal false
acceptance rate.  Thus, in cases of only partial data compromise
two-factor designs are secure to linkage attacks and can be revoked.

One way to view these results is from the perspective of reusability.
A set of access-control systems can be thought of as a series of
re-enrollments established after successive data compromise.  If any
one element -- but not both -- of the stored data $\bfS_i$ and key
$\bfK_i$ are lost, the user can simply destroy the other and
regenerate a fresh $(\bfS_{i+1},\bfK_{i+1})$ pair.  The previous,
partially compromised, enrollments do not cause any privacy leakage nor
do they enhance the adversary's ability to attack the newly enrolled
system.

Furthermore, from Theorem~\ref{thm:MultiSAR2} we learn that, in
general, the effectiveness of linkage attacks depends on the joint
structure of the error-correcting codes deployed.  Furthermore, Examples
\ref{ex.B}--\ref{ex.D} in particular give hints as to how the
collection of systems can be jointly designed to mitigate the amount
of privacy leakage or minimize the successful attack rate when some
systems have been fully compromised.  We observe from the examples
that there is a natural tradeoff between robustness to privacy leakage
and robustness to authentication attacks.  Linear dependence between
parity check matrices results in an increased probability of
successful attack while linear independence results in increased
privacy leakage. We now present a design formulation that formalizes
this tradeoff.

Our objective is to design $u$ parity-check matrices $\bfH_1, \ldots,
\bfH_u$, all full-rank $m \times n$ matrices to optimize certain
properties. To define these properties we consider all ${u}\choose{L}$
cardinality-$L$ subsets of the parity-check matrices.  Denote the
$l^\text{th}$ such subset as ${\cal S}_l$ for $1 \leq l \leq
{{u}\choose{L}}$.  The parameter $L$ corresponds to the number of
biometric systems that the adversary \emph{can potentially
  compromise}, and the subset ${\cal S}_l$ represents one set of
systems that adversary may have compromised.  For any subset ${\cal
  S}_l$, we define ${\cal S}_l[i]$ to be the index of the
$i^\text{th}$ parity check matrix in the subset.  That is $1 \leq i
\leq L$ and $1 \leq {\cal S}_l[i] \leq u$.  Further (with some abuse
of notation) we define $\bfH_{{\cal S}_l}$ to be the $L m \times n$
matrix formed by ``stacking'' all matrices in the subset into a single
matrix, i.e.,
\begin{equation*}
\bfH_{{\cal S}_l} = [\bfH^T_{{\cal S}_l[1]} \ldots
  \bfH^T_{{\cal S}_l[L]}]^T, 1 \leq l \leq {{u}\choose{L}}.
\end{equation*}
We use $r_l$ to denote the {\em collective rank} of the $l^{\text{th}}$ stacked
matrix defined as
\begin{equation*}
r_l = \mbox{rank} (\bfH_{{\cal S}_l}).
\end{equation*}
The collective rank is bounded by $1 \leq r_l \leq \min\{ lm, n\}$ and
is the privacy leakage when the adversary has gained access both to the
key and to the stored data of the $L$ systems in ${\cal S}_l$.
Theorem~\ref{thm:MultInfoLeak} establishes that for noiseless
enrollments, the stacked rank is exactly equal to the privacy leakage,
and that for noisy enrollments, the stacked rank provides an upper
bound on the privacy leakage.

Now, for each subset ${\cal S}_l$ and system $j \in \{1, \ldots, u\}$, 
define the {\em residual rank} of matrix $\bfH_j$ as
\begin{equation*}
t_{l,j} = \mbox{rank}(\bfH_{{\cal S}_l}, \bfH_j) - r_l.
\end{equation*}
Note that $t_{l,j} = 0$ if the row-space of $\bfH_j$ is spanned by the
rows of $\bfH_{{\cal S}_l}$, which would happen automatically if $j
\in {\cal S}_l$.  Also, $0 \leq t_{l,j} \leq m$, with equality to $m$
if all rows of $\bfH_j$ are linearly independent of the rows of
$\bfH_{{\cal S}_l}$.  The residual rank parameter provides a loose
characterization of the systems' linkage attack resistance to
authentication attacks.  Consider an adversary that has compromised
the keys and stored data of the enrollments of the systems in ${\cal
  S}_l$.  When $t_{l,j} = m$, the adversary does not benefit from a
higher probability of successful attack for system $j$.  On the other
hand, when $t_{l,j} = 0$ and the key of system-$j$ is compromised, the
adversary will be able to falsely authenticate at system $j$ with
probability one if the enrollments are noiseless, and with high
probability even if the enrollments are noisy.  For intermediate
values of $t_{l,j}$, determining the corresponding linkage resistance
against authentication attacks is complicated as was discussed in
Example~\ref{ex.D} of Section~\ref{sec:SARMultiple}.  Thus the
parameter $t_{l,j}$ is a rough measure of linkage attack resistance.
However, for noiseless enrollments, $t_{l,j}$ provides a lower bound
on the corresponding SAR given by
\[
P_{SA}(j,\cV) \geq 2^{-t_{l,j}},
\]
where $\cV$ are the keys and stored data for the systems in ${\cal
  S}_l$. This is because uniformly sampling from one of the
$2^{t_{l,j}}$ cosets containing the enrolled biometric is always a
strategy that is available to the attacker.

When designing a collection of systems, roughly speaking, minimizing
$r_l$ corresponds to reducing privacy leakage while maximizing
$t_{l,j}$ corresponds to reducing the probability of successful
attack.  The system designer must not only choose matrices with
desirable error-correcting properties but also consider the
optimization of these parameters across different values of $l$, $j$,
and $L$.  One possible approach is to use the following pessimistic performance
measures, $r_{\max}$ and $t_{\min}$, which are respectively defined as
\begin{align*}
%t_l & := \min_{j \in {\cal S}_l^c} t_{l,j}, \\
r_{\max} & := \max_{1 \leq l \leq {{u}\choose{L}}} r_l, \\
t_{\min} & := \min_{1 \leq l \leq {{u}\choose{L}}}\,\,\min_{j \in {\cal S}_l^c} t_{l,j},% \\
\end{align*}
where we note that the optimizing $l$ may not be the same for both
measures.  The design of a set of parity check matrices that yield low
FRRs, while minimizing $r_{\max}$ and maximizing $t_{\min}$ appears to
be a challenging avenue for future research.

\section{Conclusions}
\label{sec:conclusions}

In this paper, we presented a generalized framework for modeling
secure biometric systems and characterizing their security and privacy
properties. We conducted a detailed information-theoretic analysis of
two related types of systems based on linear error correcting codes,
namely secure sketch and fuzzy commitment.  We also considered two
variants of each scheme: keyless and keyed.  The second is a
two-factor scheme in which the biometric system is augmented by a
secret key held on a smart card. We showed that secure sketch and
fuzzy-commitment systems are equivalent in terms of the false
rejection rate, false acceptance rate, successful attack rate, and
privacy leakage during partial or full compromise of biometric
templates and smart-card keys.  We did, however, find a difference in
their storage requirements with secure sketch requiring less storage.

In either keyless or two-factor schemes, compromising the stored data
renders the biometric system vulnerable to attack. If the data stored
on the device is lost, an adversary can gain access to the system with
probability one.  However, for a two-factor system the user's
biometric sample remains protected (the information-theoretic privacy
of the user is maintained) so long as the secret key is not
compromised. In this scenario, the enrollment can be revoked and a new
one established. If, however, both the stored data and the key are
compromised, the two-factor scheme is no worse than a keyless scheme.

We also analyzed the information leakage and authentication
performance when a user's biometric is enrolled at several access
control devices.  We studied the repercussions of data compromise in a
subset of the systems.  For two-factor schemes, the successful attack
rate is no larger than the nominal false acceptance rate of the system
so long as no single system suffers from a theft of \emph{both} the
stored data and smart card key. Furthermore, no information is leaked
about the user's biometric in this case.

When some subset of systems is fully compromised, i.e., both the
stored data and the secret key are compromised, we showed that the
information leaked about the user's biometric depends on the rank of a
matrix formed by stacking the parity check matrices of the compromised
devices. The successful attack rate in this scenario depends on the
design of the parity check matrices of the compromised devices,
specifically on the number of independent rows in these matrices. We
showed via examples that, while designing multiple biometric systems,
there exists a fundamental tradeoff between the user's privacy, i.e.,
the information leaked about the underlying biometric, and the user's
security, i.e., the probability that the adversary can falsely
authenticate as a genuine user.

Many interesting problems remain open. Most importantly, in our
opinion, is the situation of multiple fully-compromised systems.
Providing the complete characterization of the tradeoff between
privacy leakage and probability of successful attack in this setting
is elusive. Such a characterization would provide guidelines for the
design of the parity check matrices for the constituent systems.  Even
with such a characterization, the joint design of parity check
matrices to achieve a point on that optimum tradeoff curve will be a 
challenge.

\bibliographystyle{styles/IEEEtran}
\bibliography{references}

\appendices
\renewcommand{\theequation}{\thesection.\arabic{equation}}
\setcounter{equation}{0}

%%%%%%%%%%%%%%%%%%%%%%%%%%%%%%%%%%%%%%%%%%%%%%%%%%%%%%%%%%%%%%%%%%%%%%
\section{Proof of Lemma~\ref{lemm:linInd}}
\label{app:lemm:linInd}

By Bayes' Theorem
\begin{align*}
\Pr[\bfH \bfA = \bfs | \tilde{\bfH} \bfA = \tilde{\bfs}] & = \frac{\Pr[\bfH
    \bfA = \bfs, \tilde{\bfH} \bfA = \tilde{\bfs}]}{\Pr[\tilde{\bfH} \bfA = \tilde{\bfs}]} \\ 
         & = \frac{\Pr\left[\left[ \!\! \begin{array}{c} \bfH
        \\ \tilde{\bfH}\end{array} \!\! \right] \bfA = \left[ \!\! \begin{array}{c}
        \bfs \\ \tilde{\bfs} \end{array} \!\! \right]\right]} {\Pr[\tilde{\bfH} \bfA
    = \tilde{\bfs}]}.
\end{align*}
Since $\tilde{\bfH}$ is full rank, all $2^{\tilde{m}}$
length-$\tilde{m}$ possible syndrome vectors $\tilde{\bfs}$ are
reachable by different choices of $\bfA$.  Also, by the theorem of
Lagrange, all cosets are of equal size.  Thus, since all realizations
of $\bfA$ are equally likely, $\tilde{\bfH} \bfA$ is uniformly
distributed, i.e., $\Pr[\tilde{\bfH} \bfA = \tilde{\bfs}] =
2^{-\tilde{m}}$.  Since $\bfH$ and $\tilde{\bfH}$ are linearly
independent, the matrix $[\bfH^T \tilde{\bfH}^T]$ has full rank.
Therefore, by the same logic as before, the numerator is equal to
$2^{-(m+\tilde{m})}$.

\section{Proof of Theorem~\ref{thm:AccuracyAnalysis}}
\label{app:thm:AcuracyAnalysis}

In the paragraphs preceding the statement of
Theorem~\ref{thm:AccuracyAnalysis}, it was proved that the FRR and FAR of
both keyed and keyless variants of fuzzy commitment and secure sketch
are the same.

%%%%%%%%%%%%%%%%%%%%%%%%%%%%%%%%%%%%%%%%%
\noindent (i) The FRR is given by
\[
P_{FR} = \Pr \big[ d(\hat{\mathbf{W}}) > \tau n \big],
\]
where, since for the legitimate user $\mathbf{D} = \mathbf{B}$ and
$\mathbf{L} = \mathbf{K}$,
\[
\hat{\mathbf{W}} = \mathop{\arg \min}_{\mathbf{W}:
\mathbf{H}\mathbf{W} = \mathbf{H}(\mathbf{A} \oplus \mathbf{B})}
d(\mathbf{W}).
\]
The FRR can be bounded by
\begin{align*}
P_{FR} &= \Pr \big[ d(\hat{\mathbf{W}}) > \tau n , \hat{\mathbf{W}} =
\mathbf{A} \oplus \mathbf{B} \big] \\
& \quad + \Pr \big[ d(\hat{\mathbf{W}}) > \tau n , \hat{\mathbf{W}}
\neq \mathbf{A} \oplus \mathbf{B} \big] \\
&\leq \Pr \big[ d(\mathbf{A} \oplus \mathbf{B}) > \tau n \big] + \Pr
\big[ \hat{\mathbf{W}} \neq \mathbf{A} \oplus \mathbf{B} \big].
\end{align*}
The decoding procedure to produce $\hat{\mathbf{W}}$ is operationally
equivalent to the optimal syndrome decoding of $\mathbf{A}$ from the
noisy version $\mathbf{B}$, since
\begin{align*}
\hat{\mathbf{W}} &= \mathop{\arg \min}_{\mathbf{W}:
\mathbf{H}\mathbf{W} = \mathbf{H}(\mathbf{A} \oplus \mathbf{B})}
d(\mathbf{W}) \\
&= \mathbf{B} \oplus \mathop{\arg \min}_{\mathbf{A'}:
\mathbf{H}\mathbf{A'} = \mathbf{H}\mathbf{A}} d(\mathbf{A'} \oplus
\mathbf{B}).
\end{align*}
Thus, the probability that $\hat{\mathbf{W}}$ fails to recover
$\mathbf{A} \oplus \mathbf{B}$ is equal to the probability of decoding
error of the code, which is bounded by
\[
\Pr \big[ \hat{\mathbf{W}} \neq \mathbf{A} \oplus \mathbf{B} \big]
\leq 2^{-n E(R) + o(n)}.
\]
The probability that $\mathbf{A} \oplus \mathbf{B}$ fails the
threshold test can be bounded by the Chernoff-Hoeffding bound
\cite{Hoeffding-JoASA-1963},
\[
\Pr \big[ d(\mathbf{A} \oplus \mathbf{B}) > \tau n \big] \leq 2^{-n
D(\tau\|p)}.
\]
Combining these two bounds yields the bound on the FRR.

%%%%%%%%%%%%%%%%%%%%%%%%%%%%%%%%

\noindent (ii) As discussed in the paragraphs preceding the statement
of Theorem~\ref{thm:AccuracyAnalysis}, in both the keyed and keyless
variants of both fuzzy commitment and secure sketch systems,
regardless of the distribution of $(\mathbf{C},\mathbf{J})$, the
syndrome is i.i.d.\ Bernoulli($0.5$). Since $\mathbf{H}$ has full row
rank, this implies that all syndromes, or equivalently all cosets, are
equally likely to be selected with probability $2^{-m}$ (there are
$2^m$ cosets). Since $P_{FA}$ is equal to the probability of selecting
a coset whose coset-leader (the minimum Hamming weight word in the
coset) has a Hamming weight not more than $\tau n$ and the number of
such cosets is not more than the total number of sequences in
$\{0,1\}^n$ with Hamming weight less than $\tau n$, it follows that
\begin{align*}
P_{FA} &\leq 2^{-m} |\{ \mathbf{w} : d(\mathbf{w}) \leq \tau n \}| \\
&= 2^{-m}\sum_{i=0}^{\tau n} |\{ \mathbf{w} : d(\mathbf{w}) = i \}| \\
&= 2^{-m} \sum_{i=0}^{\tau n} \binom{n}{i} \\
&\leq 2^{-m} 2^{n h_b(\tau)} \\
&= 2^{-n \left( \frac{m}{n} - h_b(\tau) \right) },
\end{align*}
where second inequality above is due to \cite[Lemma 8,
Ch. 10]{MacWilliams+Sloane-1977} since $\tau < 0.5$.

\noindent (iii) In both of the two-factor systems, an adversary with
knowledge of only $\mathbf{K}$ submits attack vectors
$(\mathbf{C},\mathbf{J})$ that are independent of $\mathbf{A}$. Hence,
the distribution of the syndrome is Bernoulli($0.5$), as in the
FAR analysis, and thus
\[
P_{SA}(\mathbf{K}) = P_{FA}.
\]
%

%%%%%%%%%%%%%%%%%%%%%%%%%%%%%%%%%%%%%%%

\noindent (iv) An adversary with knowledge of only $\mathbf{A}$,
submits attack vectors $(\mathbf{C},\mathbf{J})$ that are independent
of $\mathbf{K}$. Hence again the distribution of the syndrome is still
Bernoulli($0.5$), and thus
\[
P_{SA}(\mathbf{A}) = P_{FA}.
\]
%

%%%%%%%%%%%%%%%%%%%%%%%%%%%%%%

\noindent (v) Recall that $q_{_{SS}}(\mathbf{D}, \mathbf{L},
\mathbf{S}) = \mathbf{H}\mathbf{D} \oplus \mathbf{L} \oplus
\mathbf{S}$ and $q_{_{FC}}(\mathbf{D}, \mathbf{L}, \mathbf{S}) =
\mathbf{H}(\mathbf{D} \oplus \mathbf{L} \oplus \mathbf{S}) $. With
knowledge of $\mathbf{S}$, an adversary can choose $\mathbf{C} =
\mathbf{0}$ and $\mathbf{J} = \mathbf{S}$ to make
$q_{_{SS}}(\mathbf{D}, \mathbf{L}, \mathbf{S}) = q_{_{FC}}(\mathbf{D},
\mathbf{L}, \mathbf{S}) = \mathbf{0}$ so that $\hat{\mathbf{W}} =
\mathbf{0}$ and system authenticates the adversary. Thus,
\[
P_{SA}(\mathbf{S}) = 1.
\]
Since $P_{SA}(\mathcal{V}_1,\mathcal{V}_2) \geq
P_{SA}(\mathcal{V}_1)$, we also have
\[
P_{SA}(\mathbf{S}) 
= P_{SA}(\mathbf{A},\mathbf{S}) 
= P_{SA}(\mathbf{K},\mathbf{S}) 
= P_{SA}(\mathbf{A},\mathbf{K},\mathbf{S})=1.
\]
In a similar manner, one can show that with knowledge of both
$\mathbf{A}$ and $\mathbf{K}$, an adversary can set the syndrome to
any desired value and thus,
\[
P_{SA}(\mathbf{A},\mathbf{K}) = 1.
\]
%

%%%%%%%%%%%%%%%%%%%%%%%%%%%%%%%%%%%%%%%%%%

\noindent (vi) As in the proof of part (v), in the keyless versions of
the fuzzy commitment and secure sketch systems, an adversary with
knowledge of $\mathbf{S}$ alone or $\mathbf{A}$ alone can set the
syndrome to a value that makes the decoder select a coset with a
low-weight sequence with probability one. Hence,
\[
P_{SA}(\mathbf{S}) = P_{SA}(\mathbf{A}) = 1.
\]
Finally, since $P_{SA}(\mathcal{V}_1,\mathcal{V}_2) \geq
P_{SA}(\mathcal{V}_1)$, we also have
\[
P_{SA}(\mathbf{A},\mathbf{S}) = 1.
\]

% Old Appendix 1
%%%%%%%%%%%%%%%%%%%%%%%%%%%%%%%%%%%%%%%%%%%%%%%%%%%%%%%%%%%%%%%%%%%%%%

\section{Proof of Theorem~\ref{thm:SARgivenS}}
\label{app:thm:SARgivenS}

\noindent (i) Let $\mathcal{S}_a \subset \mathcal{S}$ denote the subset for which there exist $\mathbf{D},\mathbf{L}$ such that $g(\mathbf{D},\mathbf{L},\mathbf{S}) = 1$.
If $\mathbf{S} \notin \mathcal{S}_a$, then $\hat{\theta}=0$.
Therefore, the FRR must be bounded by
\[
P_{FR} \geq \Pr \big[ \mathbf{S} \notin \mathcal{S}_a \big].
\]
Since the adversary can gain access (with probability one) when $\mathbf{S} \in \mathcal{S}_a$, the SAR can be bounded as
\[
P_{SA}(\mathbf{S}) \geq \Pr \big[ \mathbf{S} \in \mathcal{S}_a \big] \geq 1 - P_{FR}. 
\]

\noindent (ii) If $\mathcal{S}_a = \mathcal{S}$, then the adversary
can always choose $\mathbf{C}$ and $\mathbf{J}$ such that
$\hat{\theta} = g(\mathbf{D},\mathbf{L},\mathbf{S}) = 1$ in order to
gain access with probability one.

%%%%%%%%%%%%%%%%%%%%%%%%%%%%%%%%%%%%%%%%%%%%%%%%%%%%%%%%%%%%%%%%%%%%%%

\section{Proof of Theorem~\ref{thm:InfoLeakage}}
\label{app:thm:InfoLeakage}

In the two-factor fuzzy commitment scheme, $\mathbf{A}$, $\mathbf{K}$,
and $\mathbf{Z}$ are mutually independent Bernoulli($0.5$) sequences
and $\mathbf{S} = \mathbf{A} \oplus \mathbf{G}^\mathrm{T}\mathbf{Z}
\oplus \mathbf{K}$. In the two-factor secure sketch scheme,
$\mathbf{A}$ and $\mathbf{K}$ are mutually independent
Bernoulli($0.5$) sequences and $\mathbf{S} = \mathbf{H} \mathbf{A} \oplus
\mathbf{K}$. Thus for both two-factor schemes, $\mathbf{A}$ and
$\mathbf{K}$ are mutually independent and so are $\mathbf{A}$ and 
$\mathbf{S}$. This implies that
\[
I(\mathbf{A};\mathbf{S}) = I(\mathbf{A};\mathbf{K}) = 0
\]
for both two-factor fuzzy and two-factor secure sketch schemes.

For the two-factor fuzzy commitment scheme,
\begin{align*}
%
%I(\mathbf{A};\mathbf{K}) &= 0, \\
%
%I(\mathbf{A};\mathbf{S}) &= 0, \\
%
I(\mathbf{A};\mathbf{S},\mathbf{K}) &= H(\mathbf{S},\mathbf{K}) -
H(\mathbf{S},\mathbf{K} | \mathbf{A}) \\
&= H(\mathbf{K}) + H(\mathbf{S}|\mathbf{K}) 
      - H(\mathbf{K}|\mathbf{A}) -
H(\mathbf{S}|\mathbf{K},\mathbf{A}) \\
&= H(\mathbf{K}) + H(\mathbf{A} \oplus \mathbf{G}^T\mathbf{Z}) 
 - H(\mathbf{K}) - H(\mathbf{G}^T\mathbf{Z}) \\
&= n - k = m.
\end{align*}
For the two-factor secure sketch scheme,
\begin{align*}
%
%I(\mathbf{A};\mathbf{K}) &= 0, \\
%%
%I(\mathbf{A};\mathbf{S}) &= H(\mathbf{S}) - H(\mathbf{S} |
%\mathbf{A}), \\
%%
%&= H(\mathbf{HA \oplus K}) - H(\mathbf{HA \oplus K} | \mathbf{A} ) \\
%%
%&= 0 \\
%%
I(\mathbf{A};\mathbf{S,K}) &= H(\mathbf{S,K}) - H(\mathbf{S,K} |
\mathbf{A}) \\
&= H(\mathbf{K}) + H(\mathbf{S}|\mathbf{K}) - H(\mathbf{K} |
\mathbf{A}) - H(\mathbf{S} | \mathbf{A, K}) \\
&= H(\mathbf{K}) + H(\mathbf{H}\mathbf{A}) - H(\mathbf{K}) - 0 \\
&= H(\mathbf{H}\mathbf{A}) = m.
\end{align*}
In the keyless fuzzy commitment scheme,
\begin{align*}
I(\mathbf{A};\mathbf{S}) &= H(\mathbf{S}) - H(\mathbf{S} | \mathbf{A})
\\
&= H(\mathbf{A} \oplus \mathbf{G}^T\mathbf{Z}) - H(\mathbf{A} \oplus
\mathbf{G}^T\mathbf{Z} | \mathbf{A}) \\
&= H(\mathbf{A}) - H(\mathbf{G}^T\mathbf{Z}) \\
&= n - k = m.
\end{align*}
And finally, in the keyless secure sketch scheme,
\[
I(\mathbf{A};\mathbf{S}) = H(\mathbf{S}) - H(\mathbf{S} | \mathbf{A})= H(\mathbf{S}) = m.
\]

% Old Appendix C

%%%%%%%%%%%%%%%%%%%%%%%%%%%%%%%%%%%%%%%%%%%%%%%%%%%%%%%%%%%%%%%%%%%%%%

\section{Proof of Theorem~\ref{thm:MultInfoLeak}}
\label{app:thm:MultInfoLeak}

To yield the main result, we show that
\begin{align*}
& I(\mathbf{A}_0 ; \mathcal{V}_1,\ldots,\mathcal{V}_u) \\
&\quad \refeq{a} I(\mathbf{A}_0 ; \mathcal{V}_1,\ldots,\mathcal{V}_l) \\
&\quad \refeq{b} I(\mathbf{A}_0 ; \mathbf{K}_1,\ldots,\mathbf{K}_l, \mathbf{H}_1\mathbf{A}_1,\ldots,\mathbf{H}_l\mathbf{A}_l) \\
&\quad \refeq{c} I(\mathbf{A}_0 ; \mathbf{H}_1\mathbf{A}_1,\ldots,\mathbf{H}_l\mathbf{A}_l).
\end{align*}
Each step is justified by the following arguments:
\begin{itemize}
\item[$(a)$] is due to the chain rule for mutual information since $(\mathcal{V}_{l+1},\ldots,\mathcal{V}_u) \Perp (\mathbf{A}_0, \mathcal{V}_1,\ldots,\mathcal{V}_l)$.
\item[$(b)$] since $\mathcal{V}_1,\ldots,\mathcal{V}_l$ is informationally 
equivalent to $\mathbf{K}_1,\ldots,\mathbf{K}_l, \mathbf{H}_1\mathbf{A}_1,\ldots,\mathbf{H}_l\mathbf{A}_l$. 
For the secure sketch system, the equivalence is immediate since 
for $i \in \{1,\ldots,l\}$, $\mathcal{V}_i = (\mathbf{S}_i,\mathbf{K}_i) = (\mathbf{H}_i\mathbf{A}_i,\mathbf{K}_i)$. 
To show the information equivalence for the fuzzy commitment system, note that 
for $i \in \{1,\ldots,l\}$, $\mathcal{V}_i = (\mathbf{A}_i \oplus \mathbf{G}_i^\mathrm{T}\mathbf{Z}_i,\mathbf{K}_i)$. 
Since $(\mathbf{H}_i\mathbf{A}_i,\mathbf{K}_i)$ is a function of $\mathcal{V}_i$, the information processing 
inequality gives $I(\mathbf{A}_0 ; \mathcal{V}_1,\ldots,\mathcal{V}_l) \geq I(\mathbf{A}_0 ; \mathbf{K}_1,
\ldots,\mathbf{K}_l, \mathbf{H}_1\mathbf{A}_1,\ldots,\mathbf{H}_l\mathbf{A}_l)$. But the information 
processing inequality also gives $I(\mathbf{A}_0 ; \mathcal{V}_1,\ldots,\mathcal{V}_l) \leq 
I(\mathbf{A}_0; \mathbf{K}_1,\ldots,\mathbf{K}_l, \mathbf{H}_1\mathbf{A}_1,\ldots,\mathbf{H}_l\mathbf{A}_l)$ 
since $\mathbf{A}_0 - (\mathbf{K}_1,\ldots,\mathbf{K}_l, \mathbf{H}_1\mathbf{A}_1,\ldots,\mathbf{H}_l\mathbf{A}_l) - (\mathbf{K}_1,\ldots,\mathbf{K}_l,\mathbf{A}_1 \oplus \mathbf{G}_1^\mathrm{T}\mathbf{Z}_1,\ldots,\mathbf{A}_l 
\oplus \mathbf{G}_l^\mathrm{T}\mathbf{Z}_l)$ forms a Markov chain. This is because 
$\mathbf{A}_i \oplus \mathbf{G}_i^\mathrm{T}\mathbf{Z}_i$ is a codeword that is independently 
chosen from the coset corresponding to $\mathbf{H}_i\mathbf{A}_i$.
\item[$(c)$] is due to the chain rule for mutual information since $(\mathbf{K}_1,\ldots,\mathbf{K}_l) \Perp (\mathbf{A}_0, \mathbf{H}_1\mathbf{A}_1,\ldots,\mathbf{H}_l\mathbf{A}_l)$.
\end{itemize}

To prove parts (i) and (ii) of Theorem~\ref{thm:MultInfoLeak}, we continue as
\begin{align*}
& I(\mathbf{A}_0 ; \mathbf{H}_1\mathbf{A}_1,\ldots,\mathbf{H}_l\mathbf{A}_l) \\
& \quad \leq I(\mathbf{A}_0 ; \mathbf{H}_1\mathbf{A}_0,\ldots,\mathbf{H}_l\mathbf{A}_0) \\
& \quad = H(\mathbf{H}_1\mathbf{A}_0,\ldots,\mathbf{H}_l\mathbf{A}_0) - H( \mathbf{H}_1\mathbf{A}_0,\ldots,\mathbf{H}_l\mathbf{A}_0 | \mathbf{A}_0 ) \\
& \quad = H(\mathbf{H}_1\mathbf{A}_0,\ldots,\mathbf{H}_l\mathbf{A}_0) \\
& \quad = \mathrm{rank}(\bfH_1,\ldots,\bfH_l).
\end{align*}
The inequality is due to the information processing inequality and the fact that $\bfA_0 - \mathbf{H}_1\mathbf{A}_0,\ldots,\mathbf{H}_j\mathbf{A}_0 - \mathbf{H}_1\mathbf{A}_1,\ldots,\mathbf{H}_l\mathbf{A}_l$ forms a Markov chain. This inequality holds with equality for noiseless enrollments. The last equality follows from 
Lemma~\ref{lemm:linInd}.

% &\quad \refeq{d} I(\mathbf{A}_0, \mathbf{H}_1\mathbf{A}_0,\ldots,\mathbf{H}_j\mathbf{A}_0 ; \mathbf{H}_1\mathbf{A}_1,\ldots,\mathbf{H}_j\mathbf{A}_j) \\
% &\quad \refeq{e} I(\mathbf{H}_1\mathbf{A}_0,\ldots,\mathbf{H}_j\mathbf{A}_0 ; \mathbf{H}_1\mathbf{A}_1,\ldots,\mathbf{H}_j\mathbf{A}_j),

% \item[$(d)$] since $\mathbf{H}_1\mathbf{A}_0,\ldots,\mathbf{H}_j\mathbf{A}_0$ is a deterministic function of $\mathbf{A}_0$.
% \item[$(e)$] is due to the chain rule for mutual information since for each $i = 1,\ldots,j$, $\mathbf{H}_i\mathbf{A}_i = \mathbf{H}_i\mathbf{A}_0 \oplus \mathbf{H}_i\mathbf{W}_i$ so that $\mathbf{A}_0 - (\mathbf{H}_1\mathbf{A}_0,\ldots,\mathbf{H}_j\mathbf{A}_0) - \mathbf{H}_1\mathbf{A}_1,\ldots,\mathbf{H}_j\mathbf{A}_j $ forms a Markov chain.

%%%%%%%%%%%%%%%%%%%%%%%%%%%%%%%%%%%%%%%%%%%%%%%%%%%%%%%%%%%%%%%%%%%%%%

\section{Proof of Theorem~\ref{thm:MultiSAR1}}
\label{app:thm:MultiSAR1}

\begin{itemize}
\item[(i)] This is an immediate corollary of
  Theorem~\ref{thm:AccuracyAnalysis}(v): Similar to the single system
  case, knowledge of $\mathbf{S}_j$ or $(\mathbf{A}_j,\mathbf{K}_j)$ -- 
  which is sufficient to generate $\mathbf{S}_j$ -- allows the adversary
  to authenticate with probability one.
%
%\item[(ii)] The adversary can set his attack vectors $\mathbf{C}$ to
%$\mathbf{A}_0$ and $\mathbf{J}$ to $\mathbf{K}_j$.  The authentication
%attack succeeds with probability at least as good as $1 - P_{FR}(j)$
%since, for $0 \leq p_j <0.5$ and $0 \leq \alpha < 0.5$, the noise
%level between $\mathbf{A}_j$ and $\mathbf{A}_0$ can only be lower than
%the noise level between $\mathbf{A}_j$ and a legitimate probe
%biometric $\mathbf{B}$.
%
\item[(ii)] The adversary can set the attack vectors $\mathbf{C}$ to
  $\mathbf{A}_i$ and $\mathbf{J}$ to $\mathbf{K}_j$.  The
  authentication attack succeeds with probability at least as large as
  $1 - P_{FR}(j)$ since, for $0 \leq p_i \leq p_j < 0.5$ and $0 \leq
  \alpha < 0.5$, the noise level between $\mathbf{A}_j$ and
  $\mathbf{A}_i$ can only be lower than the noise level between
  $\mathbf{A}_j$ and a legitimate probe biometric $\mathbf{B}$.
\item[(iii)] The compromised data $\mathcal{V}$ is independent of
  $\mathbf{S}_j$ since it does not contain $\mathbf{K}_j$, which is
  independent and i.i.d Bernoulli($0.5$).  Hence, any attack vectors
  $\mathbf{C}$ and $\mathbf{J}$ would be independent of $\mathbf{S}_j$
  and result in a uniformly distributed decoding syndrome
  $q_{SS}(\mathbf{S}_j,\mathbf{C},\mathbf{J})$ according to
  (\ref{eq:decodingsyndromeSS}).  This results in a probability of
  successful attack equivalent to the probability of false accept.
\item[(iv)] Similar to part (iii) above, the compromised data $\mathcal{V}$ is
  independent of $\mathbf{A}_j$, which is itself i.i.d
  Bernoulli($0.5$).  Hence, any attack vectors $\mathbf{C}$ and
  $\mathbf{J}$ result in a uniformly distributed decoding syndrome
  $q_{SS}(\mathbf{S}_j,\mathbf{C},\mathbf{J})$ and a probability of
  successful attack equal to the probability of false accept.
\end{itemize}

%%%%%%%%%%%%%%%%%%%%%%%%%%%%%%%%%%%%%%%%%%%%%%%%%%%%%%%%%%%%%%%%%%%%%%

\section{Proof of Theorem~\ref{thm:MultiSAR2}}
\label{app:thm:MultiSAR2}

\begin{itemize}
\item[(i)] For $i \leq l$, since both $\mathbf{S}_i$ and
  $\mathbf{K}_i$ are compromised, the syndrome $\mathbf{H}_i
  \mathbf{A}_i$ is known to the adversary.  In the case of noiseless
  enrollments, $\mathbf{H}_i \mathbf{A}_i = \mathbf{H}_i
  \mathbf{A}_0$. The linearly dependent rows of $\mathbf{H}_j$ allow
  $\mathbf{H}_j \mathbf{A}_0$ to be determined as a function of the
  compromised data.  The stored data for system $j$ can be recovered
  as $\mathbf{S}_j = \mathbf{K}_j \oplus \mathbf{H}_j \mathbf{A}_j =
  \mathbf{K}_j \oplus \mathbf{H}_j \mathbf{A}_0$. By
  Theorem~\ref{thm:MultiSAR1}(i), the adversary can therefore falsely
  authenticate with system $j$ with probability one.
\item[(ii)] Since $\mathrm{rank}(\bfH_1, \ldots, \bfH_l, \bfH_j) =
  \mathrm{rank}(\bfH_1, \ldots, \bfH_l)$, each row of $\bfH_j$ can be
  expressed as a linear combination of the rows of $\{\bfH_1, \ldots,
  \bfH_l\}$. Let $\mathbf{H}_j = \bfM_j[{\mathbf{H}}_1^T \ldots
    {\mathbf{H}}_l^T]^T$ where $\bfM_j$ is an $m_j \times (m_1 +
  \ldots + m_l)$ matrix of coefficients.  Suppose that the attacker
  chooses the attack vector pair ($\mathbf{C},\mathbf{J}$), 
  cf. Fig.~\ref{fig:framework}, such that $\mathbf{H}_j \mathbf{C} =
  \bfM_j[({\mathbf{H}}_1\mathbf{A}_1)^T \ldots
    ({\mathbf{H}}_l\mathbf{A}_l)^T]^T$ (this can always be done) and
  $\mathbf{J} = \mathbf{K}_j$. Then, the
  syndrome formed in the authentication (decoding) step of the $j$-th
  two-factor secure sketch system would be
 \begin{eqnarray*}
\mathbf{H}_j \mathbf{C} &+& \mathbf{H}_j \mathbf{A}_j \\
&=& \mathbf{H}_j
\mathbf{C} + \bfM_j[({\mathbf{H}}_1\mathbf{A}_j)^T
  \ldots ({\mathbf{H}}_l\mathbf{A}_j)^T]^T\\
&=&
\bfM_j[({\mathbf{H}}_1(\mathbf{A}_1 + \mathbf{A}_j))^T \ldots
  ({\mathbf{H}}_l(\mathbf{A}_l + \mathbf{A}_j))^T]^T.
\end{eqnarray*}
If instead $\mathbf{D} = \mathbf{B}_j$, where $\mathbf{B}_j$ is a
legitimate probe vector for system-$j$, and $\mathbf{L} =
\mathbf{K}_j$, then the syndrome formed in the authentication
(decoding) step of the $j$-th two-factor secure sketch system would be
\[
\mathbf{H}_j(\mathbf{B}_j + \mathbf{A}_j) \!=\!
\bfM_j[({\mathbf{H}}_1(\mathbf{B}_j + \mathbf{A}_j))^T ...\,({\mathbf{H}}_l(\mathbf{B}_j + \mathbf{A}_j))^T]^T
\]
Since for all $0 \leq i \leq u$, the enrollment channel crossover probability $p_i \leq \alpha_j$, 
the probe channel crossover probability, each
$\mathbf{A}_i$ is a less ``noisy'' version of $\mathbf{A}_0$ than
$\mathbf{B}_j$. Thus, the probability of system-$j$ rejecting the
specified attack vectors cannot be more than the probability that a legitimate
probe vector $\mathbf{B}_j$ is rejected (given by $P_{FR}(j)$). Thus the
authentication attack will succeed with a probability which is at
least $1 - P_{FR}(j)$.
\item[(iii)] When the rows of $\mathbf{H}_j$ are linearly independent,
  the syndrome $\mathbf{S}_j = \mathbf{H}_j \mathbf{A}_j$ is independent of the
  compromised data due to Lemma~\ref{lemm:linInd}. Another way of seeing
  this is to consider the authentication procedure in Section~\ref{sec:securesketch}.
  Using similar notation, the adversary seeks a $\hat{\mathbf{W}}$ such that
\[
\hat{\mathbf{W}} = \mathop{\arg \min}_{\mathbf{W}:
  \mathbf{H}_j\mathbf{W} = \mathbf{H}_j\mathbf{D} \oplus \mathbf{S}_j} d(\mathbf{W}).
\]
where the adversary synthesizes $\mathbf{D}$ as a function of $\mathbf{H}_i \mathbf{A}_i$, $i = 1,2,...,l$.
But, since the rows of $\mathbf{H}_j$ are linearly independent of the rows
of $\mathbf{H}_1, \mathbf{H}_2, ..., \mathbf{H}_l$, the decoding
syndrome $\mathbf{S}_j$ of the target system remains independent and uniformly
  distributed for any choice of $\mathbf{D}$ made by the adversary based on the
  compromised data. Hence, the probability of successful attack is no larger than the
  false acceptance rate.
\end{itemize}

\end{document}